 \documentclass[superscriptaddress,longbibliography,nofootinbib,twocolumn,notitlepage,aps,prx,10pt]{revtex4-2}

\usepackage{graphicx}
\usepackage{amssymb,amsmath,amsthm,mathrsfs,amsfonts,mathtools,comment}
\usepackage[colorlinks]{hyperref}
\usepackage{bm}
\usepackage{subfigure}
\usepackage{physics}
\usepackage{booktabs}

\usepackage{hyperref}
\usepackage{cleveref}

\usepackage{natbib}
\usepackage{orcidlink}

\newtheorem{thm}{\protect\theoremname}
\theoremstyle{plain}

\theoremstyle{plain}

\theoremstyle{plain}

\theoremstyle{plain}
\newtheorem*{lem*}{\protect\lemmaname}
\theoremstyle{plain}

\theoremstyle{plain}

\theoremstyle{definition}

\providecommand{\definitionname}{Definition}
\providecommand{\assumptionname}{Assumption}
\providecommand{\corollaryname}{Corollary}
\providecommand{\lemmaname}{Lemma}
\providecommand{\propositionname}{Proposition}
\providecommand{\remarkname}{Remark}

\providecommand{\theoremname}{Theorem}
\providecommand{\conjecturename}{Conjecture}

\makeatletter
\renewcommand*\env@matrix[1][\arraystretch]{%
        \edef\arraystretch{#1}%
        \hskip -\arraycolsep
        \let\@ifnextchar\new@ifnextchar
        \array{*\c@MaxMatrixCols c}}
\makeatother
\newcommand{\Tsinghua}{\affiliation{Department of Chemistry,
Tsinghua University, Beijing 100084, China}}

\newcommand{\rev}[1]{{{#1}}}

\renewcommand{\figurename}{Figure}
\begin{document}

\title{Dissipative ground state preparation in \textit{ab initio} electronic structure theory}

\author{Hao-En Li\orcidlink{0009-0002-2807-2826}}
\Tsinghua
\affiliation{Department of Mathematics, University of California, Berkeley, California 94720, USA}
\author{Yongtao Zhan\orcidlink{0000-0002-9314-0517}}
 \affiliation{Institute of Quantum Information and Matter, California Institute of Technology, Pasadena, California 91125, USA}
\author{Lin Lin\orcidlink{0000-0001-6860-9566}}
\email{linlin@math.berkeley.edu}
\affiliation{Department of Mathematics, University of California, Berkeley, California 94720, USA}
\affiliation{Applied Mathematics and Computational Research Division, Lawrence Berkeley National Laboratory, Berkeley, California 94720, USA }

\begin{abstract}
    Dissipative engineering is a powerful tool for quantum state preparation, and has drawn significant attention in quantum algorithms and quantum many-body physics in recent years. In this work, we introduce a novel approach using the Lindblad dynamics to efficiently prepare the ground state for general \textit{ab initio} electronic structure problems on quantum computers, without  variational parameters. These problems often involve Hamiltonians that lack geometric locality or sparsity structures, which we address by proposing two generic types of jump operators for the Lindblad dynamics. Type-I jump operators break the particle number symmetry and should be simulated in the Fock space. Type-II jump operators preserves the particle number symmetry and can be simulated more efficiently in the full configuration interaction space. For both types of jump operators, we prove that in a simplified Hartree-Fock framework, the spectral gap of our Lindbladian is lower bounded by a \emph{universal} constant. For physical observables such as energy and reduced density matrices, the convergence rate of our Lindblad dynamics with Type-I jump operators remains universal, while the convergence rate with Type-II jump operators only depends on coarse grained information such as the number of orbitals and the number of electrons. To validate our approach, we employ a Monte Carlo trajectory-based algorithm for simulating the Lindblad dynamics  for full \textit{ab initio} Hamiltonians, demonstrating its effectiveness on molecular systems amenable to exact wavefunction treatment.
\end{abstract}

\maketitle

\section*{Introduction} \label{sec:Introduction}

Quantum state preparation is a fundamental task in quantum simulation
and quantum algorithm design \cite{AlanAnthonyPeterEtAl2005,
VerstraeteWolfIgnacioCirac2009,MottaSunTanEtAl2020,albash2018adiabatic,CerezoArrasmithBabbushEtAl2021,ZhangLiYuan2022}. While eigenstates of a Hamiltonian can, in principle, be prepared using quantum phase estimation (QPE) and its variants, these algorithms themselves often require an initial state that has a significant overlap with the target state \cite{OBrienTarasinskiTerhal2019,GeTuraCirac2019,LinTong2020a,LinTong2022,WanBertaCampbell2022,Dutkiewicz2022heisenberglimited,DingLin2023,LeeLeeZhaiEtAl2023,BerryTongKhattarEtAl2024}. Dissipative state engineering offers a very different perspective on this problem. Rather than treating dissipation as a source of decoherence due to system-environment coupling, properly designed dissipative dynamics, such as those governed by the Lindblad equation, can encode a wide variety of strongly correlated states as the steady states of a dynamical process. Dissipative techniques, and state preparation techniques using mid-circuit measurements in general, have been widely employed in preparing matrix product states, ground states of stabilizer codes, spin systems, and other states exhibiting long-range entanglement \cite{KrausBuchlerDiehlEtAl2008,VerstraeteWolfIgnacioCirac2009,zhou2021symmetry,Cubitt2023,WangSnizhkoRomitoEtAl2023,LuLessaKimEtAl2022,Foss-FeigTikkuLuEtAl2023,KalinowskiMaskaraLukin2023}. There has been also growing recent interest in using Lindblad dynamics as an algorithmic tool for thermal state and ground state preparation \cite{MozgunovLidar2020,ChenBrandao2021,DingChenLin2024,Rall2023thermalstate,chen2023quantum,ChenKastoryanoGilyen2023,DingLiLin_KMS,gilyen2024quantum, DingLiLinZhang2024}. However, many applications have focused on Hamiltonians with special structures, as the dissipative terms often need to be carefully engineered based on the special properties of the Hamiltonian. In contrast, in quantum chemistry and materials science, \textit{ab initio} Hamiltonians lack specific geometric locality or sparsity, which significantly complicates the design of dissipative terms. 

In this work, we overcome this difficulty and present a novel method for using Lindblad dynamics to efficiently prepare the ground state for general \textit{ab initio} electronic structure problems on quantum computers. Our approach builds upon recent developments in quantum ground state preparation~\cite{DingChenLin2024}, which has the advantage of being applicable to both commuting and non-commuting Hamiltonians on an equal basis. Unlike Ref.~\cite{DingChenLin2024} which prepares the ground state using a single jump operator together with a coherent term, we propose two sets of Lindblad jump operators, termed Type-I and Type-II. Each set contains ${\rm poly} (L)$ jump operators ($L$ is the number of spatial orbitals), which are agnostic to chemical details and thus can readily be applied to  \emph{ab initio} Hamiltonians with unstructured and long range coefficients. The process does not involve variational parameters. Type-I Lindblad dynamics break the particle-number symmetry and must be simulated in the Fock space. In contrast, Type-II jump operators preserve the particle number, allowing for more efficient simulation (on both classical and quantum computers) in the full configuration interaction (FCI) space. 

\begin{figure*}[htbp]
  \centering
  \includegraphics[width=0.95\textwidth]{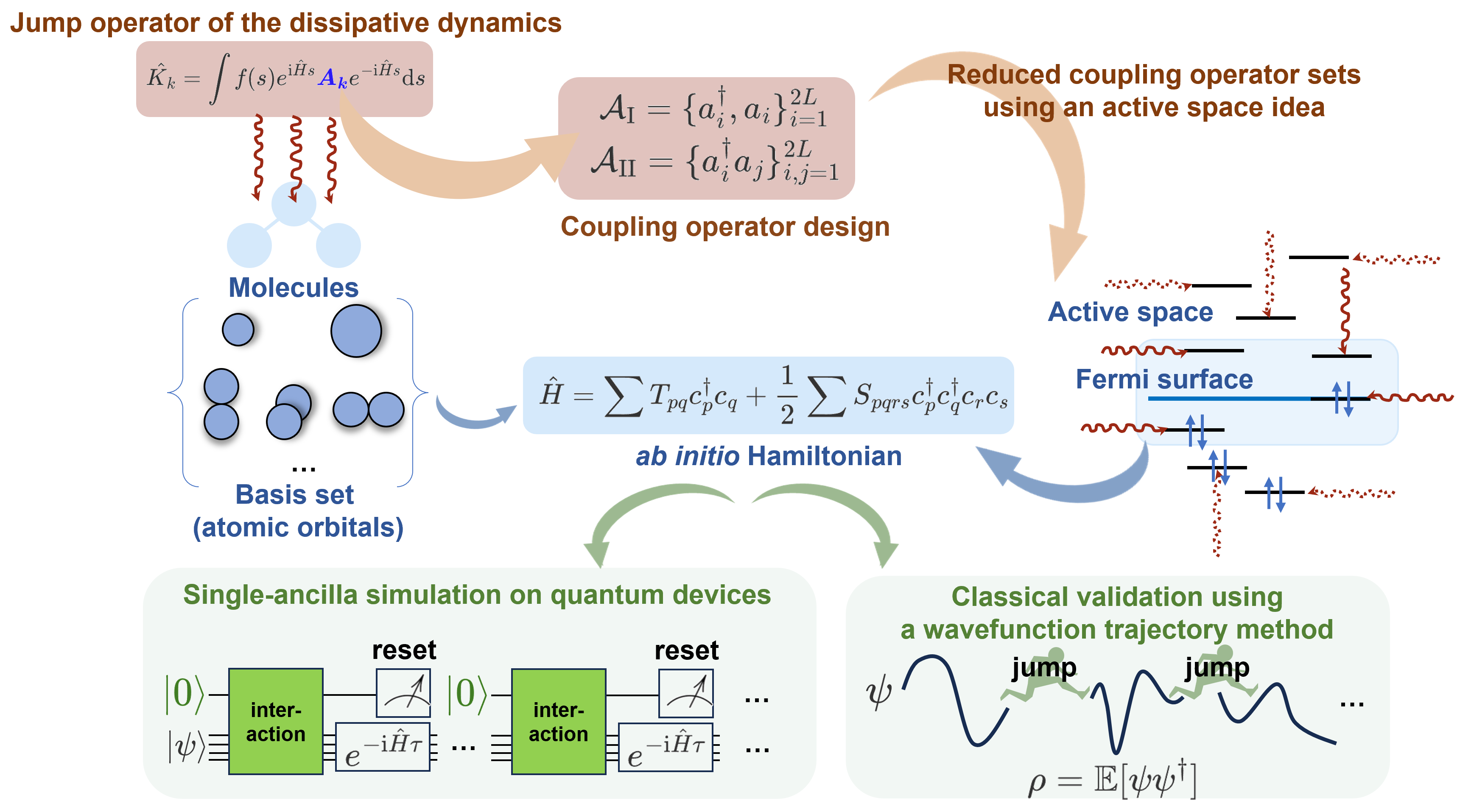}
  \caption{\rev{\textbf{Conceptual workflow illustrating the proposed dissipative ground state preparation method.} We consider the \textit{ab initio} Hamiltonian in electronic structure theory for molecular systems. The central task in this framework is to construct Lindblad jump operators, derived from either Type-I or Type-II coupling operators. An active-space strategy is employed to reduce the simulation cost. The Lindblad dynamics can be efficiently simulated on quantum devices using only single ancilla qubit, and the approach is classically validated using a wavefunction trajectory method.}}
  \label{fig:workflow}
\end{figure*}

The efficiency of Lindblad dynamics for quantum state preparation is quantified by the mixing time, which is the time required to reach the target steady state to a certain precision from an \emph{arbitrary} initial state \cite{temme2010chi,kastoryano2013quantum}. Theoretical analysis of the mixing time is in general a challenging task, and is often feasible only for specific systems, parameter regimes, or simplified settings~\cite{PhysRevLett.130.060401,kastoryano2016quantum,temme2014hypercontractivity,temme2017thermalization,BardetCapelGaoLuciaEtAl2024,alicki2009thermalization}. Our strategy is to first theoretically analyze the spectral gap of the Lindbladian, as well as dynamics of observables, such as the energy and the reduced density matrix (RDM), within a simplified Hartree-Fock (HF) framework. In this setting, the combined action of the jump operators effectively implements a classical Markov chain Monte Carlo within the molecular orbital basis.
We prove that the convergence can be provably agnostic  to specific chemical details, and in some cases, the convergence rate can be universal.
We then perform numerical simulations to examine the transferability of this behavior to the full \textit{ab initio} Hamiltonian, using an approach based on unraveled Lindblad dynamics. \rev{We also present an active-space based strategy to reduce the number of jump operators in the implementation, thereby lowering the simulation cost while preserving the convergence behavior.} This is applied to molecular systems such as $\rm BeH_2$, H$_2$O, and Cl$_2$, which are amenable to exact wavefunction treatment within the FCI space. We also apply our method to investigate the stretched square  $\rm H_4$
system, which has nearly degenerate low-energy states and poses challenges for highly accurate quantum chemistry methods such as CCSD(T).
In all cases, the Lindblad dynamics can prepare a quantum state with an energy that achieves chemical accuracy, even in the strongly correlated regime. \rev{A schematic workflow of our approach is shown in Figure~\ref{fig:workflow}.}

\section*{Results} 
\subsection*{Dissipation engineering for ground state preparation\label{subsec:gsp}}

\begin{figure*}[htbp]
    \centering
    \includegraphics[width=0.75\linewidth]{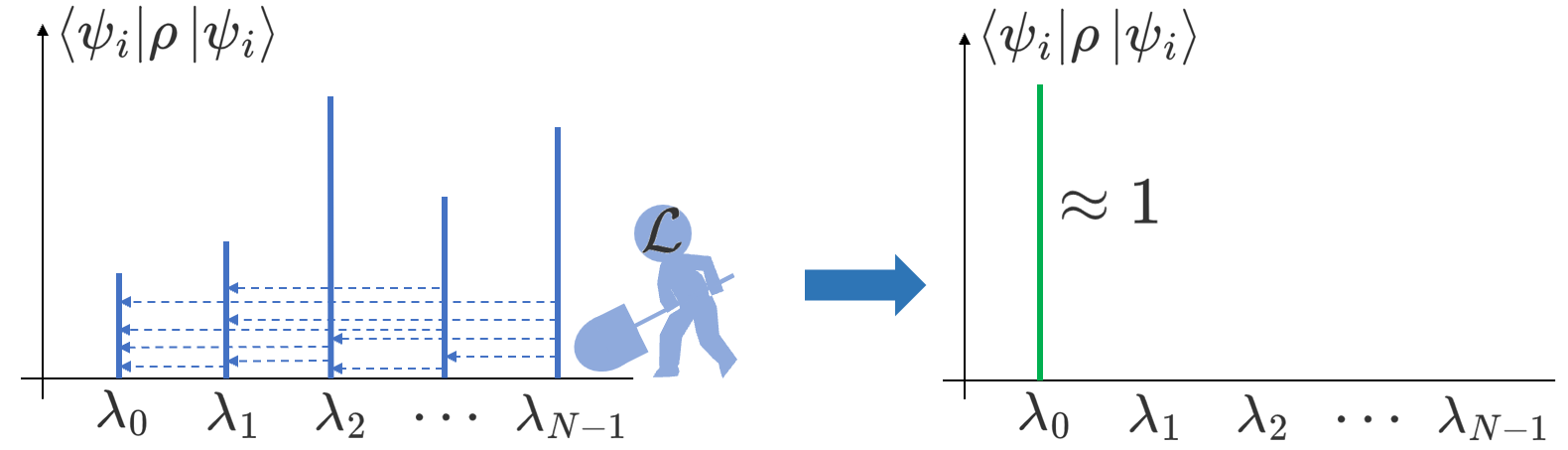}
    \caption{
\textbf{A conceptual illustration of the ``shoveling'' process in ground state preparation via Lindbladians.} The choice of jump operators ensures that the Lindbladian only allows transitions from high-energy eigenstates to low-energy eigenstates.
}
    \label{fig:shovel}
\end{figure*}

We consider Lindblad dynamics of the form:
\begin{equation}\label{eq:lind}
\begin{aligned}
  \dv{t} \rho &= \mathcal{L}[\rho] = \mathcal{L}_H[\rho] + \mathcal{L}_K[\rho] \\
  &= -\mathrm{i}[\hat{H},\rho] + \sum_k \hat{K}_k \rho \hat{K}_k^\dagger - \frac{1}{2} \{\hat{K}_k^\dagger \hat{K}_k, \rho\}.
  \end{aligned}
\end{equation}
For ground state preparation, the key object in Ref.~\cite{DingChenLin2024} is the following the jump operator:
\begin{equation}\label{eq:frequency}
  \hat{K}_k = \sum_{i,j} \hat{f}(\lambda_i - \lambda_j) \mel{\psi_i}{A_k}{\psi_j} \dyad{\psi_i}{\psi_j}.
\end{equation}
Here, $\{A_k\}$ are called (primitive) coupling operators, whose selection will be discussed in detail later. Each jump operator $\hat{K}_k$ is derived by reweighting $A_k$ in the eigenbasis $\{\ket{\psi_i}\}_{i=0}^{N-1}$ of the Hamiltonian $\hat{H}$ by  a filter function $\hat{f}(\omega)$ evaluated at the energy difference $\lambda_i - \lambda_j$.
The filter function $\hat{f}(\omega)$ is only supported on the negative axis. In other words, $\hat{f}(\lambda_i - \lambda_j) = 0$ for any $i \ge j$ (assuming $\lambda_0 < \lambda_1 \le \cdots \le \lambda_{N-1}$). As a result, the jump operator $\hat{K}_k$ only allows transitions between the eigenvectors of $\hat{H}$ that lower the energy.
Since the Lindblad dynamics generates a completely positive, trace preserving (CPTP) map \cite{Davies1974, Lindblad1976, GoriniKossakowskiSudarshan1976}, $\langle \psi_i |\rho(t) |\psi_i\rangle$ is a normalized probability distribution at time any $t$.  In the energy basis, the dynamics continuously ``shovels'' high-energy states towards low-energy ones, eventually reaching the ground state, as shown in Figure \ref{fig:shovel}.
Furthermore,  we can easily verify $\hat{K}_k \ket{\psi_0}=0$ since $\hat{f}(\lambda_j - \lambda_0) = 0$ for any $j \ge 0$. This immediately suggests that the ground state $\ket{\psi_0}\bra{\psi_0}$ is a stationary point of the Lindblad dynamics.
This dynamics is  ergodic if the ground state is the \textit{only} stationary point, and this can be achieved by a carefully chosen set of coupling operators $\{A_k\}$.

At first glance, it may seem that constructing the jump operator in Eq.~\eqref{eq:frequency} requires diagonalizing $\hat{H}$, which would clearly defeat the purpose. However, we can reformulate the definition of $\hat{K}_k$ in the time domain as 
\begin{equation}\label{eq:jump_time}
\hat{K}_k = \int_\mathbb{R} f(s) A_k(s) \, \dd s,
\end{equation}
where $A_k(s) = e^{\mathrm{i}\hat{H}s} A_k e^{-\mathrm{i}\hat{H}s}$ is the Heisenberg evolution of $A_k$, and $f(s) = \frac{1}{2\pi} \int_\mathbb{R} \hat{f}(\omega) e^{-\mathrm{i}\omega s} \, \dd \omega$ is the inverse Fourier transform of the filter function $\hat{f}$ in the frequency domain. 

In this form, the construction of the jump operator can be achieved using standard Trotter expansions for digitally simulating the Hamiltonian evolution. Additionally, the Trotter expansion can also be applied to simulate the Lindblad dynamics in Eq.~(\ref{eq:lind}). The choice of the filter function depends only on coarse-grained properties of the system, such as estimates of the spectral radius and the spectral gap of $\hat{H}$. A brief review of the selection of the filter function, the quantum simulation algorithm, \rev{as well as a rule of thumb for resource estimates are provided in the Methods section and \cref{sec:resource} in the Supplementary Information (SI) respectively}.

\subsection*{Type-I and Type-II jump operators for \textit{ab initio} calculations}

Unlike lattice problems, the second quantized Hamiltonians in \textit{ab initio} electronic structure calculations do not have clean forms such as nearest-neighbor interactions. Therefore it is important to choose simple and yet effective set of coupling operators $\{A_k\}$ that are easy to implement and allow the system to converge rapidly towards the ground state. In this work, we introduce two simple sets of primitive coupling operators, referred to as Type-I and Type-II, respectively. The corresponding jump operators can be  constructed according to Eq.~\eqref{eq:jump_time}.

We choose the set of Type-I coupling operators to be  $\mathcal{A}_{\rm I}= \{a_i^\dagger \mid i = 1,2,\cdots, 2L\}\cup \{a_i \mid i = 1,2,\cdots, 2L\}$. This includes  
all of the $4L$ (counting spatial and spin degrees of freedom) fermionic creation and annihilation operators.
Each operator can be expressed in the atomic orbital basis, molecular orbital basis, or some other basis sets. These different choices differ by a unitary matrix. Given the linear relationship between the jump operator and the coupling operator, a unitary rotation of the coupling operators will correspondingly induce a unitary rotation of the jump operators. This unitary rotation can be viewed as a gauge degree of freedom. Ideally the numerical result should be independent of this gauge. We will verify that this is indeed the case in a simplified Hartree-Fock setting.

The set of Type-II coupling operators is $\mathcal{A}_{\rm II}= \{a_i^\dagger a_j\mid i,j = 1,2,\cdots, 2L\}$ which includes every fermionic creation and annihilation pairs, and has $4L^2$ elements in total. 
Most Hamiltonians in \textit{ab initio} electronic structure calculations  are particle number preserving. Unlike Type-I coupling operators which break particle number symmetries and must be simulated in the Fock space, Type-II coupling operators  (and the corresponding jump operators) preserve particle number symmetries. The corresponding Lindblad dynamics can be simulated in the full configuration interaction (FCI) space. Note that the dimension of density matrix in the Fock formulation is $4^{2L}$, and that of the FCI space  is ${2L\choose N_{\rm e}}^2$, where $N_{\rm e}$ is the number of electrons. The difference becomes particularly significant when $N_{\rm e}$ is very small or large, such as simulating alkali metals and halogen elements in a small basis set. The particle number symmetry can be used also reduce the cost of quantum simulations, such as the Trotter error for Hamiltonian simulation $e^{-\mathrm{i}\hat Ht}$~\cite{SuHuangCampbell2021,LowSuTongEtAl2023}, or the block encoding subnormalization factors of the Hamiltonian~\cite{SuBerryWiebeEtAl2021,TongAnWiebeEtAl2021}. 

Both Type-I and Type-II sets are ``bulk'' coupling operators, meaning that dissipation is introduced on every (atomic or molecular) basis function. As will be seen below, this can be very effective in reducing the mixing time. On the other hand, this comes at the cost of introducing a large number of jump operators, \rev{which increases the} simulation cost, both on quantum computers~\cite{DingChenLin2024} and in classical simulation. We will also discuss how to reduce the number using active space ideas.

\subsection*{Universal fast convergence with Type-I set in Hartree-Fock theory}\label{sec:quasifree_theory_TypeI}

We first consider the ground state preparation via Lindbladians at the HF level before moving on to the interacting regime.  We refer readers to \rev{the Methods section} for a brief review of the Hartree-Fock theory. Essentially, after self-consistency is reached, all the information of the  Hartree-Fock theory is encoded in a non-interacting Hamiltonian \begin{equation}\hat{H} = \sum_{p,q=1}^{2L} F_{pq}a_p^\dagger a_q, 
  \end{equation}
  where the Hermitian matrix $F$ is called the Fock matrix. Let $\Phi$ be the unitary matrix that diagonalizes $F$, then the new basis set, known as molecular orbitals, is obtained by transforming the atomic orbitals using $\Phi$. For such Hamiltonians, the information contained in the many-body density operator $\rho$ is entirely stored in the one-particle reduced density matrix (1-RDM), defined as
\begin{equation}
    P_{ij} = \Tr(\rho a_j^\dagger a_i),\quad 1\le i,j\le 2L.
\end{equation}

\noindent According to the Thouless theorem \cite{Thouless1960, SzaboOstlund2012}, 
\begin{equation}
    e^{\hat{H}} a_i^\dagger = \sum_p a_p^\dagger (e^F)_{pi} e^{\hat{H}},\quad e^{\hat{H}} a_i = \sum_q  (e^{-F})_{iq} a_q e^{\hat{H}}.
\end{equation}
Therefore, we have
\begin{equation}\label{eq:kpp}
\begin{aligned}
  \hat K_{p,+} & =\int_\mathbb{R} f(s) e^{\mathrm{i} \hat H s} a_p^\dagger  e^{-\mathrm{i}\hat H s}\dd s \\
  &= \int_{\mathbb{R}}f(s) \sum_{r=1}^{2L} a_r^\dagger (e^{\mathrm{i} Fs})_{r,p} e^{\mathrm{i}\hat Hs} e^{-\mathrm{i} \hat H s}\dd s \\
  &= \sum_{r=1}^{2L} a_r^\dagger (\hat{f}(F))_{r,p}
  \end{aligned}
\end{equation}
and similarly
\begin{equation}\label{eq:kpm}
\begin{aligned}
    \hat K_{q,-}& = \int_{\mathbb{R}} f(s)e^{\mathrm{i} \hat H s} a_q  e^{-\mathrm{i}\hat H s}\dd s \\
    &= \int_{\mathbb{R}} f(s) \sum_{r=1}^{2L} (e^{-\mathrm{i} Fs})_{q,r}a_r e^{\mathrm{i} \hat Hs}e^{-\mathrm{i} \hat Hs} \dd s \\
    &= \sum_{r=1}^{2L} a_r (\hat {f}(-F))_{q,r}.
    \end{aligned}
\end{equation}
This implies that for the Type-I set, the jump operators are all linear in fermionic creation and annihilation operators. The corresponding Lindblad dynamics is quasi-free~\cite{Prosen2008,ProsenZunkovic2010,BirgerCiracGiedke2013,BarthelZhang2022}, and we can derive a closed form equation of motion for the 1-RDM 
\begin{equation}\label{eq:eom1rdm_x}
\partial_t P(t) = -\mathrm{i}[F,P(t)] + B-\frac{1}{2}[P(t)(B+C)+(B+C)P(t)].
\end{equation}
Here \begin{equation}
  \begin{aligned}
    B &= \hat f(F) \hat f(F)^\dagger = \hat f^2 (F),\\
    C &= \hat f (-F) \hat f(-F)^\dagger = \hat f^2(-F),
  \end{aligned}
\end{equation}
and we use the fact that the Fock matrix $F$ is Hermitian. A detailed derivation of Eq. (\ref{eq:eom1rdm_x}) can be found in \rev{\cref{sec:convergence} in} the SI. For any filter function $\hat f$ satisfying $\hat{f}(\omega)=1$ on $[-2\lVert{\hat H}\rVert_2,-\Delta]$ and $\hat{f}(\omega) = 0$ on $[0,+\infty)$, we have \begin{equation}
B+ C =  \hat {f}^2(F) + \hat f^2 (-F) = \mathbf{1},
\end{equation} where $\mathbf{1}$ is the identity matrix. The equation of motion Eq. (\ref{eq:eom1rdm_x}) then takes a very simple form
\begin{equation}\label{eq:1rdmbulk}
    \partial_t P(t) = -\mathrm{i} [F,P(t)] - P(t)  +\hat f(F).
\end{equation}
In particular, if we perform a unitary rotation of the fermionic creation and annihilation operators used to define the primitive coupling operators, this amounts to a gauge choice, and the final equation of motion Eq. \eqref{eq:1rdmbulk} is gauge-invariant.

From Eq. (\ref{eq:1rdmbulk}) we can easily see that $P^\star = \hat f (F)$ is the unique stationary point. In fact, let $P(t)$ and $P'(t)$ be the solution to solve Eq. (\ref{eq:1rdmbulk}) with different initial values $P(0)$ and $P'(0)$, then
\begin{equation}\label{eq:exp}
  \norm{P(t) -P'(t)}_{\rm F} = e^{-t}\norm{P(0)-P'(0)}_{\rm F},
\end{equation}
where  $\norm{A}_\mathrm{F} : =\sqrt{\Tr(A^\dagger A)}$ denotes the Frobenius norm of matrices. The detailed derivation of Eq. (\ref{eq:exp}) is provided in \rev{\cref{sec:quadbulk} in} the SI.

\begin{figure}[htbp]
    \centering
    \includegraphics[width=0.9\linewidth]{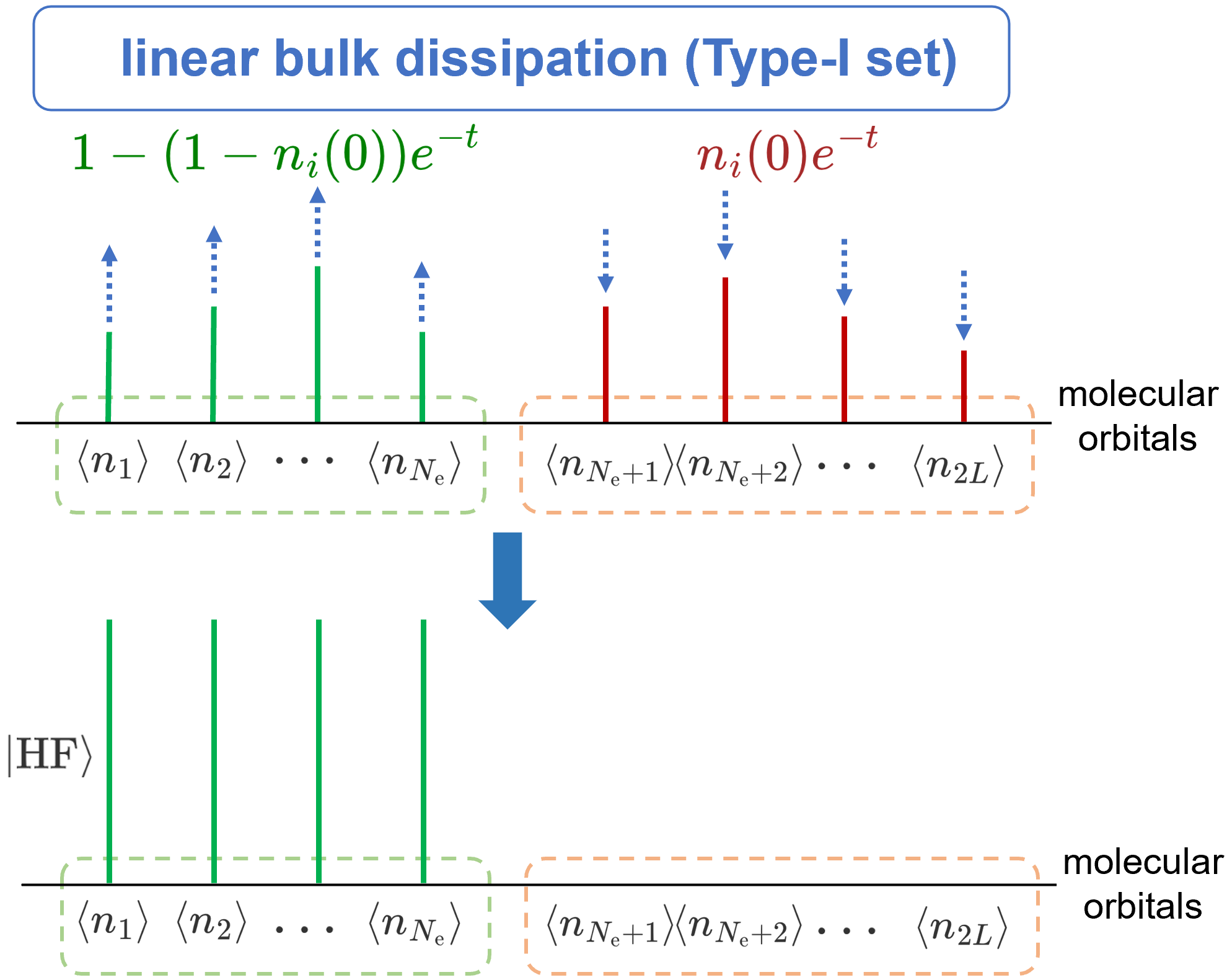}
    \caption{\textbf{A conceptual illustration of the evolution of the diagonal elements of the 1-RDM for ground state preparation with Type-I set.} The occupation numbers on each molecular orbital increase or decrease independently in an exponential rate. 
    }
    \label{fig:linearbulkconcept}
\end{figure}

We may get more insight by rewriting Eq. (\ref{eq:1rdmbulk}) in the energy basis. Specifically, we define 
\begin{equation}
    \widetilde{P} = \Phi^\dagger P \Phi =\left( \Tr(\rho c_j^\dagger c_i)\right)_{1\le i,j \le 2L},
\end{equation} where $\Phi$ is the coefficient matrix of the molecular orbitals. Then we have
\begin{equation}
\partial_t \widetilde{P}(t) = -\mathrm{i} [\Lambda , \widetilde{P}(t)] - \widetilde{P}(t) + \hat{f}(\Lambda).
\end{equation}
Here we use $F\Lambda = \Lambda \Phi$ and $\Lambda = \mathrm{diag}(\varepsilon_1,\cdots,\varepsilon_{2L})$ with $\varepsilon_1 \le\cdots \le \varepsilon_{N_{\rm e}} \le 0 < \varepsilon_{N_{\rm e}+1} \le \cdots \le \varepsilon_{2L} $. Therefore the stationary point $\widetilde{P}^\star$ is given by \begin{equation}
   \widetilde{P}^\star =  \hat f(\Lambda) = \mathrm{diag}(\underbrace{1,\cdots,1}_{N_{\rm e}},\underbrace{0,\cdots,0}_{2L-N_{\rm e}})
\end{equation}
which is consistent with the \emph{aufbau principle}, and this is achieved without explicitly diagonalizing the Fock matrix. In particular, the diagonal elements of $\widetilde{P}(t)$ evolves as follows:
\begin{equation}
    \partial_t \langle n_i\rangle  = \left(\widetilde{P}(t)\right)_{ii} = \begin{cases}-\langle n_i\rangle  + 1,\quad & i = 1,\cdots, N_{\rm e},\\
    -\langle n_i\rangle,\quad  &i=N_{\rm e}+1,\cdots,2L.
    \end{cases}
\end{equation}
    Assume the initial occupation numbers of the molecular orbitals (i.e. the diagonal elements of the initial 1-RDM $\widetilde{P}(0)$) is given by $n_i(0)$, then 
\begin{equation}
    \langle n_i\rangle = \begin{cases}
        1-(1-n_i(0))e^{-t},\quad &i=1,\cdots, N_{\rm e},\\
        n_i(0) e^{-t},\quad &i = N_{\rm e}+1,\cdots, 2L.
    \end{cases}
\end{equation}
Therefore in the energy basis, the Lindblad dynamics with Type-I jump operators can drive the occupation numbers of the lowest $N_{\rm e}$ molecular orbitals to approach $1$ exponentially, while the occupation numbers of the the remaining $2L-N_{\rm e}$ high-energy molecular orbitals exponentially approach zero (see Figure \ref{fig:linearbulkconcept}). The convergence rate is \emph{universal} and is independent of any chemical details or initial starting point. The numerical validation of this statement will be presented in the later sections.

\subsection*{Convergence with Type-II set oblivious to chemical details in Hartree-Fock theory}\label{sec:quasifree_theory_II}

Let us now carry out the calculation for Type-II jump operators in the Hartree-Fock setting. Recall $\hat{f}(\omega)=1$ on $[-2\lVert{\hat H}\rVert_2,-\Delta]$ and $\hat{f}(\omega) = 0$ on $[0,+\infty)$, using the Thouless theorem, the jump operators satisfy
\begin{equation}\label{eq:Kij}
\begin{aligned}
    \hat K_{ij}&=\int_\mathbb{R}f(s) e^{\mathrm {i} \hat H s} a_i^\dagger a_j e^{-\mathrm{i}\hat Hs}\dd s\\
    &=  \sum_{p,q=1}^{2L} \hat f(\varepsilon_p-\varepsilon_q) c_p^\dagger c_q \Phi^\ast_{ip} \Phi_{jq} = \sum_{p<q} c_p^\dagger c_q \Phi^\ast_{ip}\Phi_{jq}.
    \end{aligned}
\end{equation}
Note that $\hat{K}_{ij}$ is a quadratic operator in the fermionic operators and not Hermitian. 
This means that the Lindblad dynamics is not quasi-free, and the 1-RDM cannot satisfy a closed form equation of motion~\cite{BarthelZhang2022}.  
Despite this, we demonstrate below an explicit description of the dynamics in the energy basis. For the coherent part,
\begin{equation}
    \mathcal{L}^\dagger _H(c_r^\dagger c_s) = \mathrm{i} [\hat H, c_r^\dagger c_s] =  \mathrm{i}(\varepsilon_r - \varepsilon_s) c_r^\dagger c_s.
\end{equation}
Here $\mathcal{L}_H^\dagger$ denotes the adjoint of the superoperator $\mathcal{L}_H$ with respect to the Hilbert-Schmidt inner product. Moreover, using that $\Phi$ consists of orthonormal columns, we find 
\begin{equation}
\begin{aligned}
    \sum_{i,j=1}^{2L} \hat K_{ij }^\dagger \hat K_{ij}& = \sum_{ij}\sum_{p<q}\sum_{r<s} c_q^\dagger c_p  c_r^\dagger c_s \Phi_{ip}\Phi_{jq}^\ast \Phi_{ir}^\ast \Phi_{js}\\
    &= \sum_{p<q}c_q^\dagger  c_p c_p^\dagger c_q =\sum_{p<q}(1-n_p)n_q.
    \end{aligned}
\end{equation}
Similarly, for $r\ne s$, 
\begin{equation} 
\begin{aligned}
\mathcal{L}_K^\dagger (c_r^\dagger c_s)& = \sum_{ij} \hat K_{ij}^\dagger c_r^\dagger c_s \hat K_{ij} -\frac{1}{2}\{\hat K_{ij} ^\dagger \hat K_{ij}, c_r^\dagger c_s\}  \\
&= -\frac{1}{2}(M_r +M_s +1) c_r^\dagger c_s.
\end{aligned}
\end{equation}
Here 
\begin{equation}\label{eq:Mk}
    M_k = \sum_{p<k} c_p c_p^\dagger +\sum_{q>k} c_q^\dagger c_q = \sum_{p<k} (1-n_p) + \sum_{q>k} n_q.
\end{equation}
For $r = s$,
\begin{equation}
    \mathcal{L}_K^\dagger (c_r^\dagger c_r) =  \sum_{q>r}(1-n_r)n_q - \sum_{p<r}(1-n_p)n_r.
\end{equation}
In all the expressions above, the operators occurring in the Lindblad dynamics are all invariant to the gauge choice in the primitive coupling operators, and can all be expressed using simple operators in molecular orbitals. For a detailed derivation of the above equations, interested readers can refer to the \rev{\cref{sec:quadbulk} in the} SI.

\begin{figure}[htbp]
    \centering
    \includegraphics[width=0.9\linewidth]{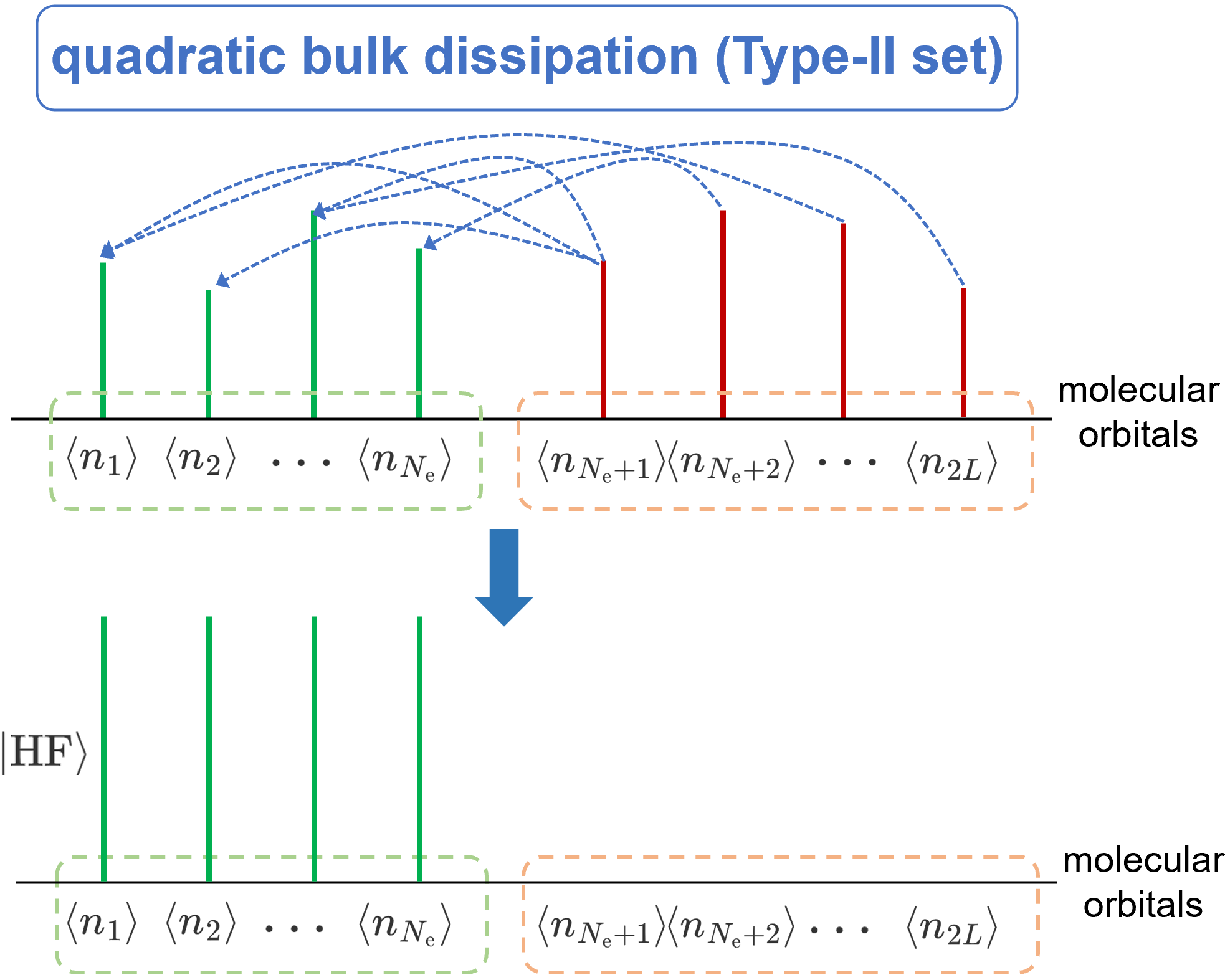}
    \caption{\textbf{A conceptual illustration of the evolution of the diagonal elements of the 1-RDM for ground state preparation with Type-II set.} It is a ``mass transport" process from higher energy orbitals to lower energy orbitals. }
    \label{fig:quadraticconcept}
\end{figure}

Consider the 1-RDM in the molecular orbital basis $\widetilde{P}= \left(\Tr(c_r^\dagger c_s\rho)\right)_{1\le s,r\le 2L} = \Phi^\dagger P \Phi$. Then the equation of motion of the entries of $\widetilde{P}$ depends on that of the 2-RDM. Specifically, for the off-diagonal elements,
\begin{equation}\label{eq:eom_quad_offdiag}
 \partial_t \widetilde{P}_{sr} =  \begin{cases}
    -\mathrm{i}(\varepsilon_s-\varepsilon_r)\widetilde P_{sr}-\frac{1}{2} \langle( M_r+M_s+1)c_r^\dagger c_s \rangle & r<s,\\
     -\mathrm{i}(\varepsilon_s-\varepsilon_r)\widetilde P_{sr}-\frac{1}{2} \langle c_r^\dagger c_s( M_r+M_s+1) \rangle & r>s.
\end{cases} 
\end{equation}
For the diagonal elements,
\begin{equation}\label{eq:eom_quad_diag}
\begin{aligned}
\partial_t \widetilde P_{rr}&=-\sum_{p<r} \widetilde{P}_{pp}+ \sum_{q>r}\widetilde{P}_{qq}  + \sum_{p<r} \langle n_pn_r\rangle-\sum_{q>r} \langle n_r n_q\rangle \\
&= -\sum_{p<r}\ev{(1-n_p)n_r} + \sum_{q>r }\ev{(1-n_r) n_q} \\
&= -\langle M_r n_r \rangle  +\sum_{q>r}\langle n_q\rangle.
\end{aligned}
\end{equation} 
Further derivations of the equation of motion for the 2-RDM will lead to the 3-RDM, and so forth. It resembles the renowned Bogoliubov--Born--Green--Kirkwood--Yvon (BBGKY) hierarchy~\cite{StefanucciVanLeeuwen2013}. To make the system solvable, we can truncate the equations by neglecting the higher-order moment terms. At this point, if we consider these matrix elements as random processes, then the equations of motion describe a \emph{classical} continuous-time Markov chain, with the stationary distribution 1-RDM approximately given by $\widetilde P^\star = \mathrm{diag}(\underbrace{1,\cdots,1}_{N_{\rm e}},\underbrace{0,\cdots,0}_{2L-N_{\rm e}})$ according to the \textit{aufbau principle}. Nonetheless, from Eq. (\ref{eq:eom_quad_offdiag}) and Eq. (\ref{eq:eom_quad_diag}), it is evident that the evolution of the RDMs are oblivious to chemical details, and are solely determined by the number of orbitals $L$ and the number of electrons $N_{\rm e}$. Moreover, the dynamics of the diagonal entries is independent of that of the off-diagonal entries. For an intuitive understanding of this, it resembles a ``mass transport" process from higher energy orbitals to lower energy orbitals. Therefore, the change in occupation number of each orbital is influenced by the electronic population of other orbitals, leading to the appearance of 2-RDM related terms in the equations (see Figure \ref{fig:quadraticconcept}). 

Given that the equations of motion of the entries of the 1-RDM are not closed, it is more challenging to analyze the convergence rate of the Type-II settings. Nonetheless, we may provide a qualitative estimation of the convergence rate using mean-field approximation. Let us first focus on the diagonal elements. For both $r>s$ and $r<s$, applying the mean-field approximation to Eq. (\ref{eq:eom_quad_offdiag}) results in the following linear homogeneous equation:
\begin{equation}
    \partial_t \widetilde{P}_{sr}  \approx \left[  -\frac{1}{2}(1+ \langle M_r+M_s\rangle) + \mathrm{i}(\varepsilon_r -\varepsilon_s)\right] \widetilde{P}_{sr}.
\end{equation}
So under the mean-field approximation, the off-diagonal entry $\widetilde{P}_{sr}$ converges exponentially to zero with an exponent of at least $\frac{1}{2}$, since $M_k$ is positive semidefinite for any $k$.

For the diagonal entries, applying the mean-field approximation on Eq. (\ref{eq:eom_quad_diag}) gives:
\begin{equation}
    \partial_t \widetilde{P}_{rr} \approx  -\langle M_r\rangle \widetilde{P}_{rr} +\sum_{q>r} \widetilde{P}_{qq}.
\end{equation}
Note that $M_k\succeq 1$ for all $r<N_{\rm e}$ and $r> N_{\rm e}+1$, the convergence rates of the off-diagonal 1-RDM entries are exponential with the exponent of at least $1$. The case $r\in \{N_{\rm e},N_{\rm e}+1\}$ is beyond the mean-field analysis. \rev{As will be shown later,} these convergence rate arguments can be verified numerically
. A detailed explanation of the results above can be found in \cref{sec:convergence} the SI. 

\subsection*{Spectral gap of Lindbladian in Hartree-Fock theory}

In this section, we provide a direct characterization of the spectral gap of the Lindbladian. \rev{The definition of the spectral gap of the Lindbladian is
\begin{equation}
\Delta_{\mathcal{L}} = - \max_{\lambda \in \text{Spec}(\mathcal L)\backslash\{0\}} \Re(\lambda).
\end{equation}
Briefly speaking, the inverse spectral gap of the Lindblad generator provides an estimate for an upper bound on the mixing time, up to logarithmic factors in the error tolerance and constants depending on the stationary state~\cite{temme2010chi, PhysRevE.92.042143}.}

Consider the following Hamiltonian, which corresponds to the dissipative part of the Lindblad dynamics,
\begin{equation}
\hat{H}_{\rm dp}=\frac12 \sum_{k} \hat{K}_k^{\dag} \hat{K}_k.
\end{equation}
Since for every jump operator $\hat{K}_k \ket{\psi_0}=0$ by construction, $\hat{H}_{\rm dp}$ can be viewed as a frustration-free\ parent Hamiltonian of the ground state $\ket{\psi_0}$. We have the following theorem (the proof is given in \cref{sec:proofthm1} in the SI).

\begin{thm}
If $[\hat{H},\hat{H}_{\rm dp}]=0$, then the spectral gap of the Lindbladian $\mathcal{L}$ is equal to the gap of $\hat{H}_{\rm dp}$.
\label{thm:gap_commute}
\end{thm}

In the Hartree-Fock setting, we have established that the equation of motion of the 1-RDM in the molecular orbital basis is independent of chemical details. Now we prove a stronger result, which states that the spectral gap of the Lindbladian is rigorously bounded from below using Type-I and Type-II jump operators in Hartree-Fock theory.

For Type-I jump operators, by calculating $\hat{H}_{\rm dp}$ in the molecular orbital basis (see \rev{\cref{sec:proofthm1}} in SI for details), we obtain
\begin{equation}
\hat{H}_{\rm dp}=\frac12 \sum_{p\le N_{\rm e}}(1-n_p) + \frac12 \sum_{q>N_{\rm e}} n_q,
\end{equation}
where $n_p,n_q$ are number operators and commute with $\hat{H}$, which is diagonal in the molecular orbital basis. Applying \cref{thm:gap_commute}, we find that the spectral gap of the Lindbladian is the same as the gap of $\hat{H}_{\rm dp}$, which is equal to $\frac12$. 

For Type-II jump operators, from previous calculations,
\begin{equation}
\hat{H}_{\rm dp}=\frac12\sum_{p<q} (1-n_p)n_q.
\end{equation}
which again only consists of number operators. Applying \cref{thm:gap_commute}, we find that the spectral gap of the Lindbladian is also equal to $\frac12$. 

For thermal state preparation, an estimate of the spectral gap provides a direct upper bound on the mixing time of the Lindblad dynamics in trace distance~\cite{temme2010chi}. This relationship, however, assumes that the stationary state is invertible, which is not satisfied by the ground state density matrix. In practice, we observe that the convergence rate of observables aligns closely with the spectral gap analysis and does not exhibit dependence on system size.

\subsection*{Numerical verification in the Hartree-Fock setting}\label{sec:numericalverfi}

\begin{figure*}
   \centering
\includegraphics[width = 0.92\textwidth]{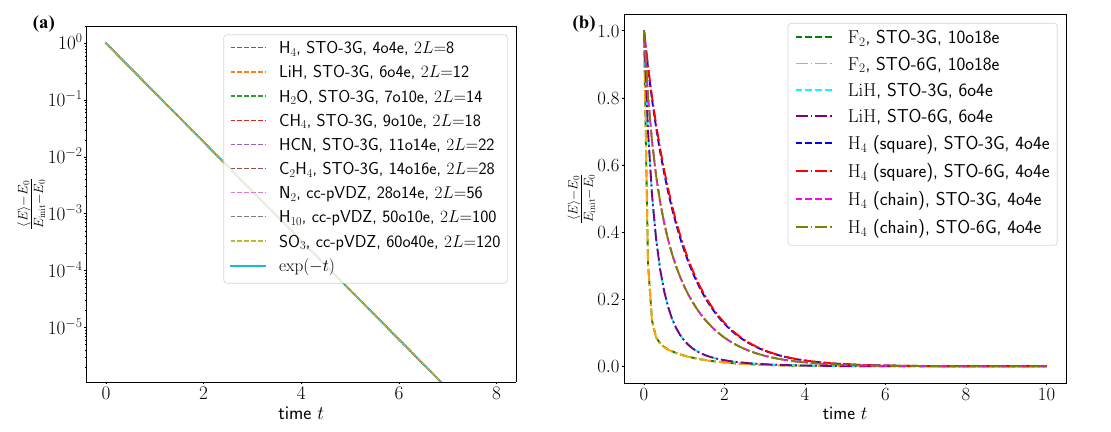}
\caption{\rev{\label{fig:oblivious}\textbf{The numerical verification of the chemical-detail-independence in the Hartree-Fock setting under the two types of dissipation.} \textbf{(a)} Convergence of energy using the Type-I set for Hartree-Fock state preparation of the molecules $\rm H_4,$ $\rm LiH$, $\rm H_2O$, $\rm CH_4$, $\rm HCN$, $\rm C_2H_4$, $\rm N_2$, $\rm H_{10}$ and $\rm SO_3$. The $y$-axis is displayed on a logarithmic scale. 
The convergence rate is universal. The dashed lines represent the convergence of energy for different molecules, while the solid green line shows the exponential decay $\exp(-t)$. \textbf{(b)} Convergence of energy using the Type-II set for $\rm F_2$, $\rm LiH$ and $\rm H_4$ molecules (square and chain geometries) in STO-3G and STO-6G. The convergence rate only depends on the number of orbitals  and the number of electrons. The dashed lines represent STO-3G results, and the dash-dotted lines represent STO-6G results, for F$_2$, LiH, and H$_4$ (square and chain geometries).}}
\end{figure*}

\begin{figure*}[htbp]
 \centering
\includegraphics[width = 0.92\textwidth]{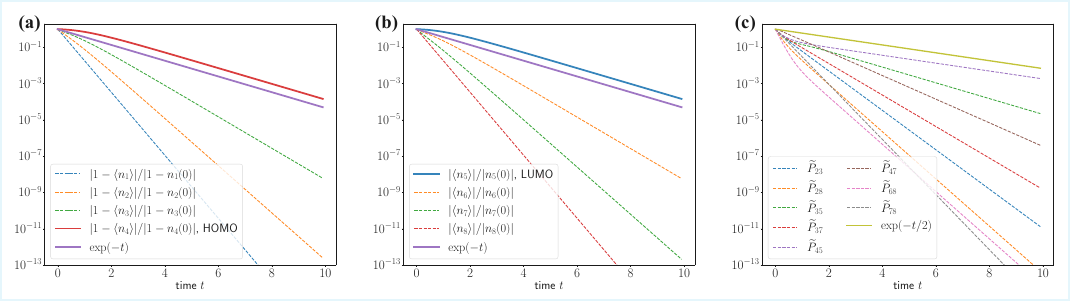}
   \caption{\label{fig:EntriesOfRDM}\rev{\textbf{Numerical demonstration of the convergence rate of the diagonal and off-diagonal entries of the 1-RDM with Type-II set.} It is tested on the $\rm H_4$  system in $\rm STO$-$\rm 3G$. \textbf{(a)} Convergence of the diagonal entries $\widetilde{P}_{rr} =\langle n_r\rangle $ of the 1-RDM for $r\le N_{\rm e}$. The colored dashed lines are diagonal entries that increase faster than the exponential reference $\exp(-t)$ (solid purple). The solid red line corresponds to the growth of the HOMO occupation number. \textbf{(b)} Convergence of the diagonal entries $\widetilde{P}_{rr} =\langle n_r\rangle $ of the 1-RDM for $r\ge N_{\rm e}+1$. The colored dashed lines are diagonal entries that decrease faster than the exponential reference $\exp(-t)$ (solid purple). The solid blue line corresponds to the decay of the LUMO occupation number. \textbf{(c)} Convergence of the off-diagonal entries of the 1-RDM. We plot $|\widetilde{P}_{ij}|/|\widetilde{P}_{ij}(0)|$ for each curve. The dashed lines represent the decay of off-diagonal entries, which all converge faster than the exponential reference $\exp(-t/2)$ shown in green solid line. }}
\end{figure*}

In this section, we numerically verify the convergence rate of 
observables such as energy and 1-RDM at the HF level. \rev{The detailed implementation of these numerical tests can be found in \cref{sec:notesOnNumerics}} in the SI.

Using the Type-I set,
Figure \ref{fig:oblivious} (a) shows that the convergence of the energy towards the ground state energy follows a universal relation $\exp(-t)$, for any molecule in any basis set. This is because the effect of the collection action of the jump operators is to \textit{independently} adjust the occupation number of each molecular orbital, until the \textit{aufbau principle} is reached. \rev{We perform each simulation by propagating the 1-RDM according to Eq. (\ref{eq:eom1rdm_x}) using the DOPRI5 solver.}

Figure \ref{fig:oblivious} (b) shows the convergence rate of the energy using the Type-II set. The test systems are $\rm F_2$ (1.4 \AA), $\rm LiH$ (1.546 \AA), chain $\rm H_4$ (2.0 {\AA}) and square $\rm H_4$ (2.0 {\AA}) using STO-3G and STO-6G basis sets. For each molecule, STO-3G and STO-6G have the same number of orbitals and electrons. The two isomers of $\rm H_4$ also share the same number of orbitals and electrons. We observe that the convergence of the systems with the same $L$ and $N_{\rm e}$ are exactly identical up to renormalization, but it varies across those molecules with different numbers, indicating a nontrivial dependence on both $L$ and $N_{\rm e}$. In each system, we perform the simulation by directly propagating the many-body density operator using DOPRI5 solver in the number preserving sector with random initialization. 

To further examine the convergence rate, we track the evolution of the diagonal and off-diagonal entries of the 1-RDM for the  $\rm H_4$ within STO-3G, as shown in Figure \ref{fig:EntriesOfRDM}. From Figure \ref{fig:EntriesOfRDM} (a) and (b), we see that the convergence rates of the diagonal entries are faster than $\exp(-t)$ except for the highest occupied molecular orbital (HOMO) and the lowest unoccupied molecular orbital (LUMO). Additionally, the off-diagonal entries exhibit convergence exponents of at least $\frac{1}{2}$. \rev{All of the numerical results are in very good agreement with the mean-field analysis discussed previously.}

\subsection*{Transferability to full \textit{ab initio} calculations}

Both Type-I and Type-II coupling operators can be readily applied in full \emph{ab initio} calculations. However, the jump operators are no longer \rev{linear or quadratic in} fermionic operators. \rev{This is because, in the \emph{ab initio} Hamiltonian
\begin{equation}
\hat{H} = \sum_{p,q=1}^{2L} T_{pq} c_p^\dagger c_q + \frac{1}{2} \sum_{p,q,r,s=1}^{2L} S_{pqrs} c_p^\dagger c_q^\dagger c_r c_s,
\end{equation}
the presence of quartic terms prevent us from applying the Thouless theorem to simplify the Heisenberg evolution of the coupling operators $A(s) = e^{\mathrm{i}\hat H s} A e^{-\mathrm{i}\hat H s}$, as in \Cref{eq:kpp,eq:kpm,eq:Kij}. Consequently, unlike in the Hartree-Fock Hamiltonian setting, we cannot expect to obtain analytic or semi-analytic solutions for the dynamics of physical observables like energy or reduced density matrices. To conceptually demonstrate the transferability of our approach to full \emph{ab initio} calculations, we therefore perform numerical simulations of the Lindblad dynamics with Type-I and Type-II jump operators in the FCI space.
}

  For small-sized systems up to $12$ spin orbitals, we may choose to propagate a many-body density operator, or a stochastic wavefunction by ``unraveling'' Lindblad dynamics and performing stochastic averages (\rev{see the Methods section as well as \cref{sec:notesOnNumerics} in SI}), in the Fock space or in the FCI space. For systems of larger sizes, the only feasible option for direct simulation is to simulate the stochastic wavefunction using Type-II jump operators in the FCI space. We quantify the convergence of the Lindblad dynamics based on the rate of energy convergence.

We note that both Type-I and Type-II sets involve a large number of jump operators, which can lead to increased simulation costs. However, in practice, the number of jump operators can be significantly reduced with minimal impact on efficiency. This is because the primary challenges in simulating chemical systems often arise in the low-energy space, particularly near the Fermi surface when we start with the Hartree-Fock initial guess. As a result, we can apply ``active space'' techniques to reduce the number of jump operators, focusing only on the most relevant degrees of freedom. For instance, 
if we start from the vacuum state, it is unnecessary to include all operators from the set $\mathcal{A}_{\rm I}=\{a_p\}_{p=1}^{2L}\cup \{a_p^\dagger\}_{p=1}^{2L}$, as the Hartree-Fock state is confined to the low-energy sector. 
Therefore, we can instead select the subset $\mathcal{S}^r_{\rm I} = \{c_{i,\uparrow}, c_{i,\uparrow}^\dagger, c_{i,\downarrow}, c_{i,\downarrow}^\dagger\}_{i= N_{\rm e}-r+1}^{N_{\rm e}+r} $, which includes only the $8r$ \rev{operators} defined around the Fermi surface under the molecular orbital basis.

\begin{figure}[htbp]
  \centering
  \includegraphics[width = 0.48\textwidth]{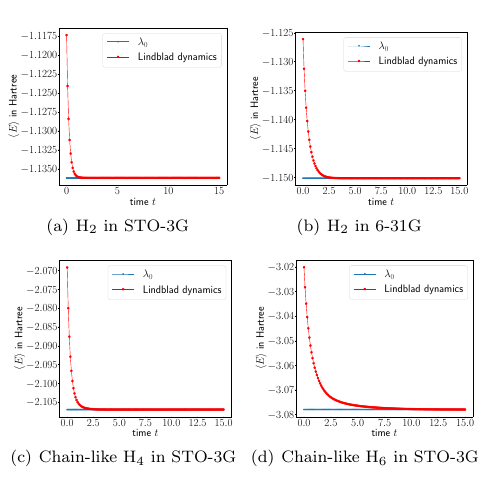}
  \caption{\label{fig:discrete}\textbf{Monte-Carlo trajectory based simulations for hydrogen chain systems.} Tested systems are: (a) $\rm H_2$ at bond length 0.7 {\AA} in STO-3G, (b) $\rm H_2$ at bond length 0.7 {\AA} in 6-31G, (c) chain-like $\rm H_4$ at bond length 0.7 {\AA} in STO-3G and (d) chain-like $\rm H_6$ at bond length 0.7 {\AA} in STO-3G. We choose $\mathcal{S}_{\rm I}^r$ as the coupling matrices with $r=1,1,1,2$ respectively and initialize with $\ket{\rm HF}$ in all of the four cases. \rev{In each panel, the red line represents the energy as a function of time, and the blue line indicates the ground-state energy of the corresponding system.}}
  \end{figure}

We perform the numerical tests for $\mathcal{S}_{\rm I}^r$. In all of the four systems demonstrated in Figure \ref{fig:discrete}, we choose the initial state to be the Hartree-Fock state. We observe that energy decreases to $\lambda_0$ with a fidelity within the chemical accuracy, which shows a good transferability of the active space reduction idea to full \emph{ab initio} calculations for the Type-I setting.

\begin{figure*}[htbp]
    \centering
    \includegraphics[width=0.80\linewidth]{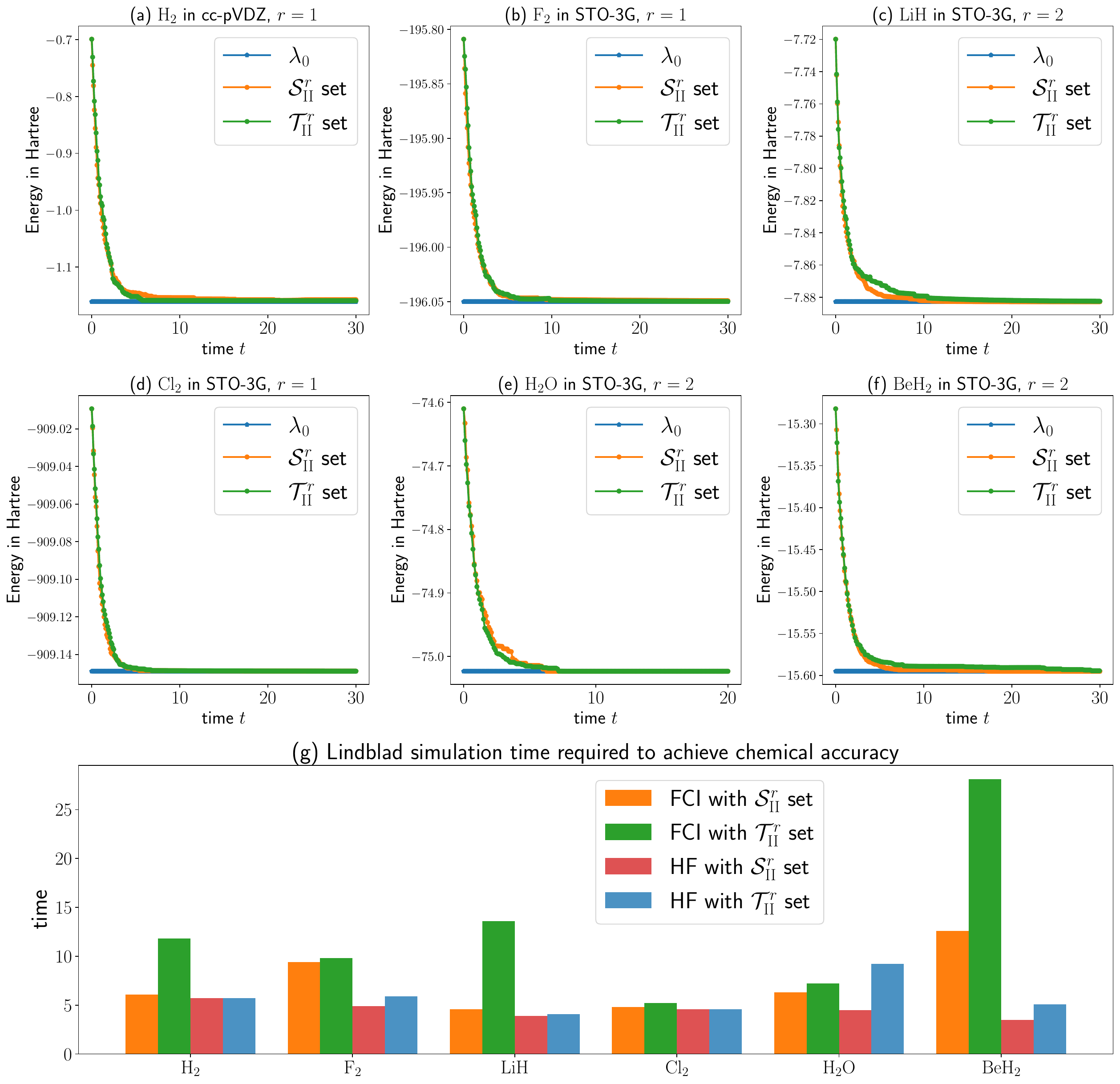}
    \caption{  \label{fig:activeII}\textbf{Monte-Carlo trajectory based simulations for full \textit{ab initio} molecular systems within the particle-number preserving sector with $\mathcal{S}^r_{\rm II}$ and $\mathcal{T}^r_{\rm II}$.} In all cases, we initialize with the (triplet) excited Hartree-Fock Slater determinant $ c_{N_{\rm e}+1,\uparrow}^\dagger c_{N_{\rm e},\downarrow}\ket{\rm HF}$.  Panels \textbf{(a-f)} display results for the $\rm H_2$, $\rm F_2$, $\rm LiH$, $\rm Cl_2$, $\rm H_2O$, and $\rm BeH_2$ molecules respectively. In each panel, the orange line and the green line represent the energy as a function of time using the $\mathcal{S}^r_{\rm II}$ and $\mathcal{T}^r_{\rm II}$ sets respectively, while the blue line indicates the ground-state energy of the corresponding system. 
    Panel \textbf{(g)} presents the Lindblad simulation time required to achieve chemical accuracy with full \textit{ab initio} and simplified Hartree-Fock Hamiltonians. The orange and green bars represent the FCI simulation time using the $\mathcal{S}^r_{\rm II}$ and $\mathcal{T}^r_{\rm II}$ sets respectively, while the red bars and blue bars represent the Hartree-Fock simulation time using the $\mathcal{S}^r_{\rm II}$ set and the $\mathcal{T}^r_{\rm II}$ set respectively.}
    \label{fig:activespace}
\end{figure*}

  \begin{figure*}[!htbp]
     \includegraphics[width = 0.97\textwidth]{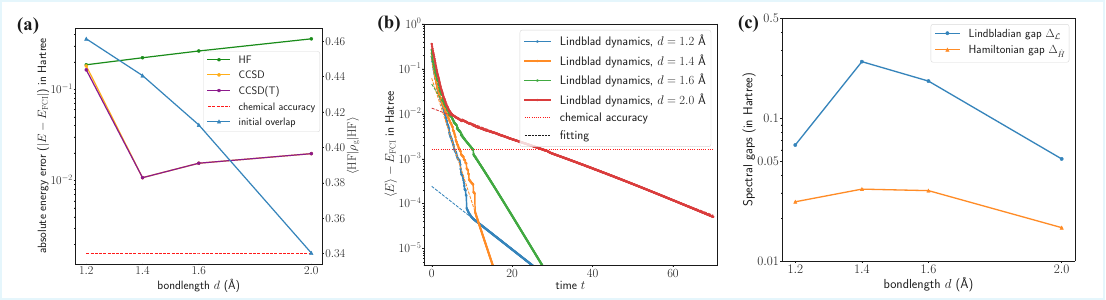}
  \caption{\label{fig:h4h6}\rev{\textbf{Monte-Carlo trajectory based simulations for the strong-correlated $\rm H_4$ molecule in STO-3G at different bond lengths.} \textbf{(a)} The accuracy of the HF, CCSD and CCSD(T) energy and the initial overlap between the HF state and the true ground state $\rho_{\rm g}$ at different bond lengths for square $\rm H_4$ system in $\rm STO$-3G. The green, yellow, and purple lines show the energy errors from HF, CCSD, and CCSD(T), respectively; the blue line shows the initial overlap, and the red dashed line marks chemical accuracy. \textbf{(b)} The energy error relative to FCI energy vs. Lindblad simulation time at bond lengths $d = 1.2, 1.4, 1.6$ and $2.0$ \AA. The solid lines (blue, orange, green, red) represent Lindblad dynamics at bond lengths $d = 1.2, 1.4, 1.6,$ and $2.0$ Å; the red dotted line indicates chemical accuracy, and the dashed lines show the fitting. \textbf{(c)} The spectral gaps of the Lindbladian $\mathcal L$ and the Hamiltonian at bond lengths $d = 1.2,1.4,1.6$ and $2.0$  \AA. The blue line denotes the Lindbladian gap $\Delta_{\mathcal{L}}$, while the orange line denotes the Hamiltonian gap $\Delta_{\hat H}$.}}
  \end{figure*}

  Similarly, for the Type-II set, we can start from the Hartree-Fock state or low excited Slater determinant and include only the jump operators defined on a small number of orbitals around the Fermi surface. In this case, we consider the following reduced set of particle-number preserving coupling operators
\begin{equation}
  \mathcal{S}_{\rm II}^r = \{c_{i,\sigma}^\dagger  c_{j,\tau}|i,j\in \{N_{\rm e}-r+1,\cdots, N_{\rm e}+r\}, \sigma,\tau \in \{\uparrow, \downarrow\}\}, 
\end{equation}
which has $16r^2$ \rev{coupling operators} in total. In practice, setting $r=1$ or $r=2$ is typically sufficient to achieve convergence of the system to its ground state. The corresponding numerical results are presented in Figure \ref{fig:activeII}. To compare the convergence rates of the full \emph{ab initio} and Hartree-Fock state preparation with the reduced set $\mathcal{S}_{\rm II}^r$, we begin from the (triplet) excited Slater determinant $ c_{N_{\rm e} +1 ,\uparrow}^\dagger c_{N_{\rm e},\downarrow}\ket{\rm HF}$. Notably, the convergence rate of the full \emph{ab initio} state preparation is observed to be not much slower than that of the Hartree-Fock method, in the sense of Lindblad simulation time required to reach chemical accuracy.

In fact, the set of Type-II orbitals can be further compressed. 
\rev{For instance, we may take the subset $\mathcal{T}_{\rm II}^r$ of $\mathcal{S}_{\rm II}^r$ that contains only the hopping between nearest energy levels of molecular orbitals:}  
\begin{equation}
  \begin{aligned}
      \mathcal{T}_{\rm II}^r &= \{c_{i,\sigma}^\dagger c_{j,\tau}|i,j\in\{N_{\rm e}-r+1,\cdots,N_{\rm e}+r\},\\
      &\quad \abs{i-j}=1,\quad \sigma,\tau \in \{\uparrow, \downarrow\}\}
      \end{aligned}
  \end{equation}
  
  The number of operators in $\mathcal{T}^r_{\rm II}$ increases only linearly with $L$ for fixed $r$. The numerical convergence behaviors with $\mathcal{T}^r_{\rm II}$ are shown in Figure \ref{fig:activeII}. We observe that the further compression of $\mathcal{S}_{\rm II}^r$ maintains the simulation efficiency, in the sense that the Lindblad simulation time required to achieve chemical accuracy with the $\mathcal{T}_{\rm II}^r$ set remains comparable to or only slightly longer than that needed with the $\mathcal{S}_{\rm II}^r$ set.

  The ground state preparation via Lindbladians with the set $\mathcal{T}^r_{\rm II}$ can perform well even in strongly correlated systems where the Hartree-Fock state poorly approximates the true ground state. 
  Examples like the stretched square $\rm H_4$ 
  highlight these challenges due to their nearly degenerate low energy states, which even highly accurate methods like CCSD(T), often referred to as 
  the ``gold standard" in molecular quantum chemistry, struggle with \cite{PaldusPiecuchPylypowEtAl1993,LeeHugginsHead-GordonEtAl2019,HugginsOGormanBryanEtAl2022}. Our results show that Lindblad dynamics effectively captures the correlation energy. As shown in Figure \ref{fig:h4h6} (a), as the bond length of the square $\rm H_4$ system increases, the energy accuracy of the Hartree-Fock initial guess decreases. Concurrently, the initial overlap between the Hartree-Fock state and the true ground state diminishes, indicating a growing extent of strong correlation. 
   However, as illustrated in Figure \ref{fig:h4h6} (b), the convergence of Lindblad dynamics remains largely unaffected by the degree of strong correlation starting from the Hartree-Fock initial state at various bond lengths. \rev{Meanwhile, we  observe that increasing the extent of strong correlation leads to slower asymptotic convergence of the Lindblad dynamics. Specifically, as shown in \Cref{fig:h4h6} (c), at bond lengths where the Hamiltonian gap $\Delta_{\hat H}$ (i.e., the gap between the ground and first excited state energies) becomes smaller, the spectral gap of the Lindbladian is also reduced. Correspondingly, the exponential fitting results in \Cref{fig:h4h6} (b) demonstrate slower \emph{asymptotic }convergence rates. These observations suggest that stronger correlations in molecular systems \emph{can} lead to slower convergence to the ground state.}

\rev{It is also noteworthy that in \Cref{fig:h4h6} (b), the behaviors at short and long bond lengths are  different. At short bond lengths, the pre-asymptotic decay is relatively rapid and the error quickly falls below chemical accuracy, making the preparation appear easier. In contrast, at long bond lengths, the pre-asymptotic decay is shorter-lived and the convergence is dominated by the smaller spectral gap, requiring a longer time to reach chemical accuracy. }

\section*{Discussion} \label{sec:Conclusion}

To our knowledge, this work is the first Lindblad-based ground state preparation algorithm for \emph{ab initio} electronic structure calculations. The Lindblad dynamics is employed as an algorithmic tool for dissipative state engineering, \rev{which can be constructed without relying on variationally adjusted parameters.
A notable advantage of this approach is that the effectiveness of the method can be nearly independent of the quality of the initial state. This stands in sharp contrast to quantum phase estimation (QPE), whose cost depends directly on the initial state's overlap with the target state and can fail if this overlap vanishes.
The ``shoveling'' process in dissipative state preparation shares some conceptual similarities with various forms of imaginary time evolution (ITE) \cite{ZhangCarlsonGubernatis1995, ZhangKrakauer2003, BoothThomAlavi2009, McArdleJonesEndoEtAl2019, MottaSunTanEtAl2020, HugginsOGormanBryanEtAl2022, LeePhamReichman2022, JiangZhangBaumgartenEtAl2024}, but also exhibits notable differences. A direct implementation of ITE through the application of $e^{-\tau \hat{H}}$ does not yield a completely positive trace-preserving (CPTP) map, and its quantum realization again requires nontrivial overlaps between the initial and target states. The quantum imaginary time evolution (QITE) algorithm~\cite{MottaSunTanEtAl2020} addresses some of these challenges, but it relies on a tomography procedure and its cost can scale explicitly exponentially with system size.
} 

In the dissipative state preparation framework, we prove that the Lindblad dynamics with Type-I and Type-II jump operators can converge rapidly in a simplified Hartree-Fock setting, and validate the transferability to full \emph{ab initio} calculations even for systems exhibiting strong correlation behaviors.
In order to perform numerical simulation for larger systems with tens to hundreds of spin-orbitals, 
 even propagating the state vector in the FCI space can be very costly, and advanced simulation methods such as quantum Monte Carlo methods or tensor network-based methods must be employed. \rev{For practical implementation on quantum devices, the simulation of Lindblad dynamics involves repeatedly applying circuit blocks with intermediate measurements, continuing until the dynamics approach a fixed point. As a result, a large mixing time in certain systems can lead to a substantial overall simulation cost. It is therefore important to develop a stronger theoretical framework for analyzing convergence rates in the ground state preparation problem. Recent progress in this direction can be found in \cite{ZhanDingHuhnEtAl2025}.}
This work may also provide new perspectives of ground state preparation in other areas, such as nuclei physics, fermions with random coefficients (e.g., the SYK model), or optimization problems on unstructured graphs. 

\section*{Methods}
\subsection*{Choice of filter function and sketch of quantum simulation algorithm}\label{sec:filter}

In this section, we discuss the choice of filter function in Ref. \cite{DingChenLin2024} and briefly review quantum algorithms for simulating the Lindblad dynamics.

We begin by reviewing the definition of the jump operator $\hat{K}_k$. Since the eigenvectors of $\hat{H}$ are typically not accessible, we express $\hat{K}_k$ in the time domain as follows:
\begin{equation}
  \hat{K}_k = \int_\mathbb{R} f(s) A_k(s) \dd s
\end{equation}
where $A_k(s) = e^{\mathrm{i}\hat Hs} A_k e^{-\mathrm{i}\hat Hs}  $ is the Heisenberg evolution of $A_k$ and $f(s)=\frac{1}{2\pi }\int_{\mathbb{R}} \hat{f}(\omega) e^{-\mathrm{i}\omega s}\dd \omega$ is the inverse Fourier transform of the filter $\hat f$ in the frequency domain. One possible choice for the filter function in the frequency domain $\hat f$ is given by the following form \cite{DingChenLin2024}
\begin{equation}\label{eq:filter}
  \hat{f}(\omega) := \frac{\mathrm{erf}(\frac{\omega+a}{\delta_a})-\mathrm{erf}(\frac{\omega+b}{\delta_b})}{2}
\end{equation}
where $\mathrm{erf}(\omega):= \int_{0}^{\omega}\frac{2}{\sqrt{\pi}} e^{-t^2}\dd t$ denotes the error function. Here $a$ is chosen to be an energy cutoff satisfying $a > 2\lVert\hat H\rVert_2$ and $b$ is chosen to be the spectral gap of the Hamiltonian $\Delta := \lambda_1-\lambda_0$. The parameters $\delta_a$ and $\delta_b$ are chosen to be on the same order of $a$ and $b$ respectively. In this setup, $\hat{f}$ is approximately supported on the interval  $[-2\lVert{\hat{H}}\rVert_2,-\Delta]$ (Note that the largest eigenvalue difference is $\abs{\lambda_i-\lambda_j}\le 2\lVert{\hat{H}}\rVert_2$). The inverse Fourier transform $f$ can also be computed analytically as  
\begin{equation}
  f(s) = \frac{1}{2\pi \mathrm{i}s} \left(\exp(\mathrm{i}as-\delta_a^2s^2/4)-\exp(\mathrm{i}bs-\delta_b^2s^2/4)\right)
\end{equation}
where $f(0) = \frac{a-b}{2\pi}$ is obtained by taking the limit $s\to 0$. $f(s)$ is a smooth complex-valued function with the modulus $\abs{f(s)}$ exhibiting a rapid decay when $|s|\to \infty$. Specifically, $f(s)$ approximately vanishes when $|s|>S_s$ for some $S_s = \Theta(1/\Delta)$, which allows us to truncate the infinite integral and use the trapezoidal quadrature rule to approximate the jump operator $\hat K_k$:
\begin{equation}\label{eq:quadrature}
  \hat{K}_k\approx \int_{-S_s}^{S_s} f(s) A(s) \dd s \approx\sum_{l = -M_s}^{M_s} f(s_l) A(s_l) w_l
\end{equation}
where $M_s$ is the number of quadrature nodes on $[0, S_s]$ or $[-S_s, 0]$, $s_l = l\Delta s$ and $\Delta s = S_s/M_s$. The weights $\{w_l\}$ are chosen to be $\Delta s/2$ for $l=\pm M_s$ and $\Delta s$ for $-M_s+1\le l \le M_s-1$.

To simulate the Lindblad dynamics $\exp(\mathcal{L} t )$ on quantum devices, we can begin with a first order Trotter splitting $\exp(\mathcal L t)\approx (\exp(\mathcal L_H\tau)\cdot \exp(\mathcal L_K\tau))^{t/\tau}$. The coherent dynamics $\exp({\mathcal{L} _H\tau})$ is just the Hamiltonian simulation $\exp(-\mathrm{i}\hat H \tau)$. For the nonunitary dissipative part $\exp(\mathcal{L}_K \tau)$, we can reduce this problem to a dilated Hamiltonian simulation up to a partial trace on the ancilla qubit, namely
\begin{equation}
  \exp({\mathcal{L}_K}\tau)[\rho] = \Tr_{a} e^{-\mathrm{i}\widetilde{K}\sqrt{\tau}}(\dyad{0_a}\otimes \rho) e^{\mathrm{i}\widetilde{K}\sqrt{\tau}} + O(\tau^2).
\end{equation}
Here, $\Tr_a$ denotes the partial trace operation on the ancilla qubit. For simplicity we consider only one jump operator $K,$ and $\widetilde{K}$ is defined by the Hermitian dilated matrix $\widetilde{K}=\smqty(0&K^\dagger \\ K&0)$. Then the dilated Hamiltonian simulation $e^{\mathrm{i}\widetilde{K}\sqrt{\tau}}$ can be efficiently performed on a quantum computer using a second order Trotter splitting according to the discretized time evolution of $K$ shown in Eq. (\ref{eq:quadrature}) (see \cite[Section III]{DingChenLin2024} for details).

\subsection*{Review of Hartree-Fock theory in the second quantized representation}\label{sec:HF}

The HF theory finds a self-consistent single-particle operator approximation to the many-body Hamiltonian taking the form\begin{equation}\label{eq:HF}
  \hat{H}_{\rm HF} = \sum_{p,q=1}^{2L} F_{pq}a_p^\dagger a_q,\quad F_{pq} = h_{pq}+V_{pq}-K_{pq}.
\end{equation}
Here $L$ is the number of spatial orbitals, and $2L$ is the number of spin orbitals. The operators $a_p^\dagger$ and $a_p$ are fermionic creation and annihilation operators with respect to the orthonormalized spin orbitals. Here $h_{pq}$ is a fixed single-particle matrix.
The direct Coulomb and Fock exchange terms, denoted by $V_{pq}$ and $K_{pq}$, respectively, should be solved self-consistently with respect to the one-particle density matrix (1-RDM), defined as
\begin{equation}
  D_{pq}  = \sum_{i=1}^{N_{\rm e}} \Phi_{pi}\Phi_{qi}^\ast,\quad p,q=1,2,\cdots,2L.
\end{equation}
Here $\Phi\in\mathbb{C}^{2L\times 2L}$ is a unitary matrix called the molecular orbital coefficients, which are eigenfunctions of the Fock matrix $F$. We denote $F \Lambda = \Lambda \Phi$ with $\Lambda = \mathrm{diag}(\varepsilon_1,\cdots,\varepsilon_{2L})$.

Once $D$ is fixed, the HF Hamiltonian in Eq. (\ref{eq:HF}) is a quadratic Hamiltonian. Its ground state has an explicit expression in the Slater-determinant form:
\begin{equation}\label{eq:HFstate}
  \ket{\rm HF} = c_1^\dagger\cdots c_{N_{\rm e}}^\dagger \ket{\rm vac},
\end{equation}
where the new set of creation operators $\{c_i^\dagger\}$ are given by the unitary transform of $\{a_i^\dagger\}$ via 
\begin{equation}\label{eq:molecular}
  (c_1^\dagger,\cdots,c_{2L}^\dagger)  = (a_1^\dagger, \cdots, a_{2L}^\dagger) \Phi.
\end{equation}
 Note that $\ket{\rm HF}$ is the ground state of the converged Hamiltonian $\hat H_{\rm HF}$ with eigenvalue $E_0=\sum_{k=1}^{N_{\rm e}}\varepsilon_k$. This sum of eigenvalues can also be expressed as $\Tr(FD)$, which differs from the Hartree-Fock energy by a nonlinear term that depends only on $D$.

 \subsection*{Monte-Carlo trajectory based method for unraveled Lindblad dynamics}\label{sec:unravel}

To solve the Lindblad dynamics for ground state preparation at the FCI level, we may directly propagate the many-body density operator $\rho(t)\in \mathbb{C}^{2^L\times 2^L}$ using a differential equation solver \cite{Lidar2019}. Another approach is to ``unravel'' the Lindblad dynamics for the many-body density operator \cite{DalibardCastinMolmer1992,DumZollerRitsch1992}. Broadly speaking, we employ a family of Monte-Carlo-type algorithms where we only propagate the state vector $\ket{\psi(t)}$ in some stochastic schemes, and the many-body density operator $\rho(t)$ at the time $t$ can be retrieved by taking the average of the random matrix $\dyad{\psi(t)}$, i.e.
$
  \rho(t) = \mathbb{E}\dyad{\psi(t)}.
$ The deterministic Lindblad equation for the dynamics of the many-body density operator is now expressed using stochastic pure-state trajectories, thus leading to a quadratic reduction in dimensionality, 
 at the cost of incorporating statistical averaging across multiple runs.

The simplest setting is the discrete form of unraveling, or quantum jump method \cite{ChristieEastmanRomanEtAl2022}. The quantum-jump pure-state dynamics is evolved under an effective non-Hermitian Hamiltonian $\hat H - \mathrm{i}/2\sum_k \hat K_k^\dagger \hat K_k$, with stochastic quantum jumps occurring intermittently throughout the evolution. Specifically, it can be described by the following stochastic differential equation (SDE):\cite{BreuerPetruccione2002,MoodleyPetruccione2009, ChristieEastmanRomanEtAl2022}  
\begin{equation}\label{eq:poisson}
  \begin{aligned}
  \dd \psi &= \left(-\mathrm{i}\hat H-\frac{1}{2}\sum_k (\hat K_k ^\dagger \hat K_k-\langle \hat K_k ^\dagger \hat K_k\rangle)\right)\psi \dd t \\
  &+ \sum_k\left(\frac{\hat K_k}{\sqrt{\langle\hat K_k^\dagger \hat K_k\rangle}}-1\right)\psi \dd N_t^k.
  \end{aligned}
\end{equation}
Here, $N_t$ denotes a Poisson process with a splitting $N_t = \sum_k N_t^k$. For a sufficiently small time step  
$\Delta t$, the Poisson increment $\Delta N_t$ takes the values $0$ (no jump) or $1$ (jump) with expectation value $\mathbb{E}(\Delta N_t) =\sum_k\lVert{\hat K_k\psi}\rVert^2\Delta t=\sum_k\langle\hat K_k^\dagger \hat K_k\rangle\Delta t $. The Poisson processes $\{N_t^k\}$ are mutually independent with intensities given by $\lVert{\hat K_k\psi}\rVert^2=\langle\hat K_k^\dagger \hat K_k\rangle$. This implies that, in the event of a jump, we select the jump operator $\hat K_k$ to apply to $\psi$ with a probability proportional to $\langle\hat K_k^\dagger \hat K_k\rangle$ for $k=1,2,\cdots, N$. It can be shown that the density operator, defined as $\rho(t) = \mathbb{E} \dyad{\psi(t)}$ indeed solves the Lindblad dynamics, by calculating $\dv{\psi \psi^\dagger}{t}$ using It\^o's lemma for the Poisson process and then taking the expectation.

For a Monte-Carlo-type simulation of Eq. (\ref{eq:poisson}), we first discretize the time interval by the time step $\Delta t$. Then at each step, we randomly pick up $k\in [N+1]$ (assume we have $N$ jump operators  in total) with respect to the distribution 
\begin{equation}
p_k = \begin{cases}
  \lVert\hat K_k \psi_n\rVert^2 \Delta t,\quad k\le N\\
  1-\sum_{\ell =1}^N p_\ell, \quad k=N+1.
\end{cases}
\end{equation}
 If $k=1,\cdots, N$, we update the trajectory using $\psi_{n+1}= \hat K_k\psi_n/\langle \hat K_k^\dagger \hat K_k\rangle$. If $K=N+1$, we update using $\psi_{n+1}= \psi_n-(\mathrm{i}\hat H+\frac{1}{2}\sum_k \hat K_k ^\dagger \hat K_k)\psi_n\Delta t$. Then the many-body density operator $\rho(t_n)$ at time $t_n$ can be approximated by taking the average of the pure states $\dyad{\psi(t_n)}$ over the trajectories \cite{Lidar2019}.

We can also consider a variant of the Monte Carlo-type algorithm that is slightly different but essentially equivalent to the one described above. Notice that \cite{Lidar2019}
\begin{equation}
  \begin{aligned}
    &~~~~\norm{\exp(\mathrm{-i}\left(H-\frac{\mathrm{i}}{2}\sum_k \hat K_k^\dagger \hat K_k\right)\Delta t)\psi(t)}^2 \\
    &= 1-\sum_k \lVert{\hat K_k\psi(t)}\rVert ^2\Delta t +\mathcal O((\Delta t)^2)
\end{aligned}
\end{equation}
which implies that the decaying evolution governed by the non-Hermitian effective Hamiltonian primarily dictates the probability distribution. Consequently, we actually do not need to compute $\lVert \hat{K}_k \psi_n \rVert^2$ at every time step. Instead, we can first calculate $\lVert \psi_n \rVert^2$ and sample a random number
 $R\sim \mathrm{U}(0,1)$ to decide whether jump or not. If $R<\norm{\psi_n}^2$, we just propagate the trajectory using $\psi_{n+1} = \psi_n - \left(\mathrm{i}\hat H+\frac{1}{2}\sum_{k} \hat K_k^\dagger \hat K_k\right) \psi_n \Delta t$. If $R\ge \norm{\psi_n}^2$, we calculate each $\langle \hat K_k^\dagger \hat K_k\rangle$ and update $\psi_{n+1} = {\hat K_k\psi_n}/{\langle\hat K_k^\dagger \hat K_k\rangle}$ with a probability proportional to ${\langle\hat K_k^\dagger \hat K_k\rangle}$ for $k=1,2,\cdots,N$. After this, we reset the random number $R\sim \mathrm{U}(0,1)$. Essentially, this corresponds to evolving the trajectory using $\hat H-\mathrm{i}/2\sum_k \hat K_k^\dagger\hat K_k$ deterministically causing the norm of $\psi$ to decrease, 
   until a random quantum jump occurs. At this point, the norm is restored to $1$, and the process repeats \cite{BarthelZhang2022}.

In all of the steps in the unraveling algorithm, only matrix-vector multiplication is involved.

\vbox{}

\section*{Data availability}
The data that support this study are available upon request.

\vbox{}

\section*{Code availability}

The codes that support this study are available on GitHub via \url{https://github.com/haoen2021/LindbladAbInitio}.

\vbox{}

\section*{Acknowledgments} 

H.-E. L. acknowledges support from the National Natural Science Foundation of China (Grant No. 223B1011) and the Tsinghua Xuetang Talents Program for funding a summer visit to the University of California, Berkeley, and the High-Performance Computing Center of Tsinghua University for providing computational resources. 
Y. Z. acknowledges support from the Institute for Quantum Information and Matter, an NSF Physics Frontiers Center.
L. L. acknowledges support from the U.S. Department of Energy, Office of Science, through the Accelerated Research in Quantum Computing Centers program (Quantum Utility through Advanced Computational Quantum Algorithms, Grant No. DE-SC0025572), and from the Simons Investigator in Mathematics program (Grant No. 825053).
The authors thank Zhiyan Ding, Zhen Huang, and Jakob Huhn for helpful discussions.

\vbox{}

\section*{Author contributions} 
L.L. conceived the original study.  H.-E. L. and L.L. carried out theoretical analysis to support the study. H.-E. L. and Y. Z. carried out numerical calculations to support the study. All authors discussed the results of the manuscript and contributed to the writing of the manuscript.

\section*{Competing Interests}
The authors declare no competing interests.

\widetext
\section*{References}
\bibliographystyle{naturemag} 
\bibliography{ref}

   \widetext
\appendix

\newpage
 \begin{center}
    \textbf{SUPPLEMENTARY MATERIALS}
 \end{center}
 \renewcommand{\figurename}{Supplementary Figure}
\renewcommand{\tablename}{Supplementary Table}
\setcounter{figure}{0}
\section{Resource estimate}\label{sec:resource}
\rev{The main goal of this manuscript is to demonstrate that, for practical quantum chemistry applications, the mixing time can be short, making dissipative ground-state preparation via Lindblad dynamics potentially useful in practice. Conceptually, the total cost for dissipative ground-state preparation can be decomposed into three components: (1) the cost of constructing the Lindbladian, in particular the jump operators; (2) the cost of simulating the Lindblad dynamics per unit time; and (3) the total simulation time, which can be upper bounded by the mixing time $t_{\rm mix}$. The overall resource cost is the product of these three factors. The simulation algorithm described in~\cite[Section III]{DingChenLin2024} can be viewed as a first-order method for simulating continuous-time Lindblad dynamics. For a typical physical Hamiltonian (spin, fermion, etc.) defined on $N$ sites, the total resource cost to achieve precision $\epsilon$ is~\cite[Theorem 1]{DingChenLin2024}
\[
\widetilde{\mathcal{O}}\!\left(t^2_{\rm mix} \, \Delta^{-1} \, \operatorname{poly}(N) / \epsilon\right),
\]
where $\Delta$ is the spectral gap. \cite[Theorem 2]{DingChenLin2024} further presents a discrete-time algorithm that reduces the cost from quadratic to nearly linear in $t_{\rm mix}$.
Therefore, for the end-to-end cost to scale polynomially with $N$, the main technical challenge is to estimate the mixing time and establish that it grows at most polynomially with the system size.
}

\section{Solving quasi-free Lindblad dynamics}\label{sec:solvingQuasiFree}
We consider the solution of the Lindblad dynamics that is quasi-free, i.e. with Hamiltonians that are quadratic in the fermionic creation and annihilation operators and Lindblad jump operators that are linear in the fermionic creation and annihilation operators \cite{Prosen2008, ProsenZunkovic2010, BirgerCiracGiedke2013, BarthelZhang2022}. 

Specifically, we consider $\hat H =\sum_{p,q}^{2L} F_{pq} a_p^\dagger a_q$ with a Hermitian coefficient matrix $F=(F_{pq})_{1\le p,q\le 2L}$, and the following Lindblad jump operators:
\begin{equation}
    \hat K_{p,+} = \sum_{r=1}^{2L} a^\dagger_{r} K_{r,p},\quad \hat K_{q,-} = \sum_{r=1}^{2L} a_r J_{q,r}, \quad K_{r,p}, J_{q,r}\in\mathbb{C}.
\end{equation}
We will derive the equation of motion for the one-particle reduced density matrix (1-RDM) with respect to the many-body density operator $\rho$, defined as
\begin{equation}
    P_{ij} = \Tr(\rho a_j^\dagger a_i),\quad 1\le i,j\le 2L.
\end{equation}

From the Lindblad equation:
\begin{equation}
    \partial_t \rho = -\mathrm{i}[\hat{H},\rho]  +\sum_p (\hat{K}_{p,+}\rho \hat K_{p,+}^\dagger-\frac{1}{2}\{\hat K_{p,+}^\dagger \hat K_{p,+},\rho\})+\sum_q (\hat {K}_{q,-}\rho \hat K_{q,-}^\dagger-\frac{1}{2}\{\hat K_{q,-}^\dagger \hat K_{q,-},\rho\}),
\end{equation}
the equation of motion of $P_{ij}$ is given by
\begin{equation}\label{eq:eom1rdm}
    \partial_t P_{ij}(t)= -\mathrm{i}\Tr([\hat{H},\rho]a_j^\dagger a_i)+\sum_{p,\bullet\in\{+,-\}}   \Tr(\hat K_{p,\bullet}^\dagger  a_j^\dagger a_i \hat K_{p,\bullet}\rho)-\frac{1}{2}\Tr(\{\hat K_{p,\bullet}^\dagger \hat K_{p,\bullet},a_j^\dagger a_i\}\rho).
\end{equation}

According to the canonical anticommutation relations (CAR),
each term in Eq. (\ref{eq:eom1rdm}) can be simplified as follows:
\begin{equation}\label{eq:1rdmcoherent}
  -\mathrm{i}\Tr([\hat{H},\rho]a_j^\dagger a_i) = -\mathrm{i}\sum_{p=1}^{2L}\left( F_{ip} P_{pj}-  F_{pj} P_{ip}\right),
\end{equation}
\begin{equation}\label{eq:1rdmKpplus}
    \sum_{p}   \Tr(\hat K_{p,+}^\dagger  a_j^\dagger a_i \hat K_{p,+}\rho)-\frac{1}{2}\Tr(\{\hat K_{p,+}^\dagger \hat K_{p,+},a_j^\dagger a_i\}\rho)= \sum_{p}K_{jp}^\ast K_{ip}-\frac{1}{2}\sum_{p,r}(K_{jp}^\ast K_{rp} P_{ir} + K_{rp}^\ast K_{ip} P_{rj} ),
\end{equation}
and
\begin{equation}\label{eq:1rdmKpminus}
    \sum_{p}   \Tr(\hat K_{p,-}^\dagger  a_j^\dagger a_i \hat K_{p,-}\rho)-\frac{1}{2}\Tr(\{\hat K_{p,-}^\dagger \hat K_{p,-},a_j^\dagger a_i\}\rho)= -\frac{1}{2}\sum_{p,r}(J_{jp}^\ast J_{rp} P_{ir} + J_{rp}^\ast J_{ip} P_{rj}) .
\end{equation}

Next we provide detailed derivations of Eq. (\ref{eq:1rdmcoherent}) to (\ref{eq:1rdmKpminus}).

For Eq. (\ref{eq:1rdmcoherent}),
\begin{equation}
\begin{aligned}
    \Tr([\hat H,\rho]a_j^\dagger a_i) &= \sum_{p,q}\Tr( F_{pq} a_p^\dagger a_q \rho a_j^\dagger a_i)-\Tr(F_{pq}\rho a_p^\dagger a_q a_j^\dagger a_i)\\
    &=\sum_{pq}F_{pq} \Tr(\rho (a_j^\dagger a_i a_p^\dagger a_q-a_p^\dagger a_q a_j^\dagger a_i))\\
    & =\sum_{pq }F_{pq}\left(\delta_{ip}{\Tr} (\rho a_j^\dagger a_q)-\delta_{qj} {\Tr}(\rho a_p^\dagger a_i)\right) +\sum_{pq} F_{pq} \Tr(\rho(a_p^\dagger a_j^\dagger a_q a_i - a_j^\dagger a_p^\dagger a_i a_q))\\
    & = \sum_{q} F_{iq} P_{qj} -\sum_p F_{pj} P_{ip} + \sum_{pq} F_{pq} \Tr(\rho((-a_j^\dagger a_p^\dagger) (-a_ia_q) - a_j^\dagger a_p^\dagger a_i a_q))\\
    & = \sum_{q} F_{iq} P_{qj} -\sum_p F_{pj} P_{ip} .
    \end{aligned}
\end{equation}
Here we use the definition $P_{ij} = \Tr(\rho a_j^\dagger a_i)$ and the canonical anticommutation relations.

For Eq. (\ref{eq:1rdmKpplus}), we first notice that
\begin{equation}\label{seq:diss}
\begin{aligned}
    \Tr((\hat K_{p,+}^\dagger a_j^\dagger a_i \hat K_{p,+} \rho) - \frac{1}{2}\{\hat K_{p,+}^\dagger \hat K_{p,+},a_j^\dagger a_i\}\rho ) & = \frac{1}{2}\Tr([\hat K_{p,+}^\dagger ,a_j^\dagger a_i]\hat K_{p,+} \rho)+\frac{1}{2}\Tr(\rho \hat K_{p,+}^\dagger [a_j^\dagger a_i, \hat K_{p,+}]). \\
    \end{aligned}
\end{equation}

For each term in Eq. (\ref{seq:diss}), we expand $\hat K_{p,+}$ and  $\hat K_{p,+}^\dagger$, yielding
\begin{equation}
\begin{aligned}
&~~~~\Tr(([\hat K_{p,+}^\dagger ,a_j^\dagger a_i]\hat K_{p,+} + 
\hat K_{p,+}^\dagger [a_j^\dagger a_i, \hat  K_{p,+}])\rho)\\ & = \sum_{r,s=1}^{2L}\Tr(  K_{rp}^\ast K_{sp}(a_ra_j^\dagger a_i a_s^\dagger - a_j^\dagger a_i a_r a_s^\dagger +a_r a_j^\dagger a_i a_s^\dagger - a_r a_s^\dagger a_j^\dagger a_i)\rho)
        \end{aligned}
\end{equation}
Using the algebraic relations $[a_p, a_j^\dagger a_i] =\delta_{pj} a_i$, $[a_q^\dagger, a_j^\dagger a_i]=-\delta_{qi} a_j^\dagger $ (can be verified directly using CAR), we obtain:
\begin{equation}
\begin{aligned}
\Tr(([\hat K_{p,+}^\dagger ,a_j^\dagger a_i]\hat K_{p,+} + 
\hat K_{p,+}^\dagger [a_j^\dagger a_i, \hat  K_{p,+}])\rho)&=\sum_{r,s=1}^{2L}\Tr( K_{rp}^\ast K_{sp}([a_r,a_j^\dagger a_i]a_s^\dagger - a_r[a_s^\dagger , a_j^\dagger a_i])\rho)\\
    &=  \sum_{r, s=1}^{2L} \Tr(K_{rp}^\ast K_{sp}(\delta_{rj}a_ia_s^\dagger +\delta_{si}a_r a_j^\dagger)\rho)\\
    & = \sum_{s=1}^{2L} K_{j,p}^\ast K_{s,p}{\Tr}(a_i a_s^\dagger \rho ) + \sum_{r=1}^{2L} K_{r,p}^\ast K_{i,p}{\Tr}(a_ra_j^\dagger \rho).
    \end{aligned}
\end{equation}
Taking summation with respect to the index $p$ on both sides, we get
\begin{equation}
    \begin{aligned}
        &~~\sum_p \Tr(([\hat K_{p,+}^\dagger ,a_j^\dagger a_i]\hat K_{p,+} + 
\hat K_{p,+}^\dagger [a_j^\dagger a_i, \hat  K_{p,+}])\rho)\\
        & = \sum_p\left(\sum_{s=1}^{2L} K_{j,p}^\ast K_{s,p}[\delta_{is}-{\Tr}(a_s^\dagger a_i \rho )] + \sum_{r=1}^{2L} K_{r,p}^\ast K_{i,p}[\delta_{rj}-{\Tr}(a_j^\dagger a_r \rho )]\right)\\
        & = 2\sum_p K_{ip }K_{jp}^\ast  - \sum_{p,r}(K_{jp}^\ast K_{rp} P_{ir} + K_{rp}^\ast K_{ip} P_{rj} ),
    \end{aligned}
\end{equation}
which proves Eq. (\ref{eq:1rdmKpplus}). Here we use $\Tr\rho =1$. Eq. (\ref{eq:1rdmKpminus}) can be verified by following almost the same procedure.

Using Eq. (\ref{eq:1rdmcoherent}) to Eq. (\ref{eq:1rdmKpminus}), it follows readily that the equation of motion $P(t)$ has the following closed form:
\begin{equation}\label{eq:eom1rdm_xs}
\partial_t P(t) = -\mathrm{i}[F,P(t)] + B-\frac{1}{2}[P(t)(B+C)+(B+C)P(t)].
\end{equation}
The positive semidefinite matrices $B$ and $C$ are defined as
\begin{equation}\label{eq:defofBandC}
    B_{ij} = \sum_p K_{ip} K_{jp}^\ast,\quad C_{ij} = \sum_{p} J_{ip} J_{jp}^\ast.
\end{equation}

Eq. (\ref{eq:eom1rdm_xs}) is a linear autonomous system for $P(t)$ of size $2L\times 2L$. Therefore, we can solve Lindblad dynamics of the quasi-free fermionic systems of large
sizes by studying the evolution of the 1-RDM. Once we have the equation of motion for $P(t)$, we can obtain the evolution of energy expectation value (or other quadratic observables) of $\rho(t)$ using the 1-RDM:
\begin{equation}
    E(t) =\Tr(\rho (t)\hat H) = \sum_{p,q=1}^{2L}\Tr(\rho(t)a_p^\dagger a_q  F_{pq}) = \Tr(P(t)F).
\end{equation}

\section{Equations of motion of 1-RDM with Type-II jump operators}\label{sec:quadbulk}

In this section, we provide a detailed derivation of the equation of motion of the 1-RDM with the Type-II set. We consider the converged Hartree-Fock Hamiltonian $\hat H = \sum_{p,q=1}^{2L} F_{pq} a_p^\dagger a_q$ with a Hermitian coefficient matrix $F$ and the following set of coupling operators:
\begin{equation}
    \mathcal{A}_{\rm II} = \{a_i^\dagger a_j\mid i ,j = 1,2,\cdots,2L\}.
\end{equation}

Assume $F$ can be diagonalized as  $F\Lambda = \Lambda \Phi$, where
 $\Phi$ is a unitary matrix and the diagonal entries of $\Lambda  = \mathrm{diag}(\varepsilon_p)$ are ordered non-decreasingly. In the molecular orbital basis,
the HF Hamiltonian can be further simplified as a weighted summation of the number operators on the molecular orbitals, namely $\hat H = \sum_{p=1}^{2L} \varepsilon_p c_p^\dagger c_p.$
 
 Next we consider the equation of motion of the $1$-RDM under the energy basis of the HF Hamiltonian. We first calculate 
 \begin{equation}\label{eq:coh}
\begin{aligned}
    \mathcal{L}^\dagger_{H}(c_r^\dagger c_s)&=\mathrm{i}[\hat H,c_r^\dagger c_s ]= \mathrm{i}\sum_p \varepsilon_p c_p^\dagger c_p c_r^\dagger c_s- \varepsilon_p c_r^\dagger c_s  c_p^\dagger c_p\\
    & =\mathrm{i}\sum_p \delta_{rp}\varepsilon_p c_p^\dagger c_s - \varepsilon_p c_p^\dagger c_r^\dagger c_p c_s -\delta_{ps}\varepsilon_p c_r^\dagger c_p  + \varepsilon_p c_r^\dagger c_p ^\dagger c_s c_p\\
    & = \mathrm{i}\sum_p \delta_{rp} \varepsilon_p c_p^\dagger c_s -\varepsilon_p (-c_r^\dagger c_p^\dagger )(-c_sc_p) -\delta_{ps} \varepsilon_p c_r^\dagger c_p +\varepsilon_p c_r^\dagger c_p^\dagger c_s c_p\\
    & = \mathrm{i}(\varepsilon_r - \varepsilon_s) c_r^\dagger c_s .
    \end{aligned}
\end{equation}

  Next, we compute the jump operators using the Thouless theorem
 \begin{equation}
 \begin{aligned}
     \hat K_{ij} = \int_\mathbb{R} f(s)  e^{i\hat H s} a_i^\dagger a_j  e^{-i\hat H s}\dd s =  \sum_{p,q=1}^{2L}\hat f(\varepsilon_p-\varepsilon_q) c_p^\dagger c_q \Phi_{ip}^\ast \Phi_{jq} = \sum_{p<q} c_p^\dagger c_q \Phi_{ip}^\ast \Phi_{jq}.
     \end{aligned}
  \end{equation}
 Here we use $\hat{f}|_{[-2\lVert \hat H\rVert_2, -\Delta]}= 1$ and $\hat f|_{[0,+\infty)}=0.$ Therefore, using the fact that $\Phi$ consists of orthonormal columns, we have
 \begin{equation}
     \sum_{i,j=1}^{2L} \hat K_{ij}^\dagger \hat K_{ij} =\sum_{ij}\sum_{p<q}\sum_{r<s} c_q^\dagger c_p c_r^\dagger c_s \Phi_{ip} \Phi_{jq}^\ast \Phi_{ir}^\ast \Phi_{js} = \sum_{p<q} c_q^\dagger c_pc_p^\dagger c_q=\sum_{p<q} (1-n_p)n_q.
 \end{equation}

Similarly, we calculate
\begin{equation}\label{eq:crcs}
    \mathcal{L}_K^\dagger (c_r^\dagger c_s) =\sum_{i,j }\hat K_{ij}^\dagger c_r^\dagger c_s \hat K_{ij} -\frac{1}{2}\{\hat K_{ij}^\dagger \hat K_{ij} , c_r^\dagger c_s\} = \sum_{p<q} c_q^\dagger c_pc_r^\dagger c_s c_p^\dagger c_q - \frac{1}{2}\{c_q^\dagger c_p c_p^\dagger c_q, c_r^\dagger c_s\}.
\end{equation}
 To simplify Eq. (\ref{eq:crcs}), we begin with rewriting it as
\begin{equation}\label{eq:crcsrr}
    \mathcal{L}_K^\dagger(c_r^\dagger c_s) = \frac{1}{2}\sum_{p<q} (c_q^\dagger c_p c_r^\dagger c_s c_p^\dagger c_q- c_q^\dagger c_p c_p^\dagger c_q c_r^\dagger c_s)+
\frac{1}{2} \sum_{p<q} (c_q^\dagger c_p c_r^\dagger c_s c_p^\dagger c_q -c_r^\dagger c_s c_q^\dagger c_p c_p^\dagger c_q).
\end{equation}
For the first term,
\begin{equation}
\begin{aligned}
    c_q^\dagger c_p c_r^\dagger c_s c_p^\dagger c_q- c_q^\dagger c_p c_p^\dagger c_q c_r^\dagger c_s&=\delta_{ps} c_q^\dagger c_p c_r^\dagger c_q - c_q^\dagger c_p c_r^\dagger c_p^\dagger c_s c_q - \delta_{rq} c_q^\dagger c_p c_p^\dagger c_s + c_q^\dagger c_p c_p^\dagger c_r^\dagger c_q c_s\\ 
    & = \delta_{ps} c_q^\dagger c_p c_r^\dagger c_q - c_q^\dagger c_p (-c_p^\dagger c_r^\dagger) (-c_q c_s) -\delta_{rq} c_q^\dagger c_p c_p^\dagger c_s + c_q^\dagger c_p  c_p^\dagger c_r ^\dagger c_q c_s\\
    &= \delta_{ps} c_q^\dagger c_p c_r^\dagger c_q-\delta_{rq} c_q^\dagger c_p c_p^\dagger c_s.
    \end{aligned}
\end{equation}
Likewise, for the second term,
\begin{equation}\label{eq:secondterm}
c_q^\dagger c_p c_r^\dagger c_s c_p^\dagger c_q -c_r^\dagger c_s c_q^\dagger c_p c_p^\dagger c_q =\delta_{pr} c_q^\dagger c_s c_p^\dagger c_q -\delta_{sq} c_r^\dagger c_p c_p^\dagger c_q.
\end{equation}
So we have,
\begin{equation}
\begin{aligned}
    \mathcal{L}_K^\dagger (c_r^\dagger c_s) &=\frac{1}{2}\left(\sum_{q>s} c_q^\dagger c_s c_r^\dagger c_q -\sum_{p<r} c_r^\dagger c_p c_p^\dagger c_s + \sum_{q>r} c_q^\dagger c_s c_r^\dagger c_q -\sum_{p<s} c_r^\dagger c_p c_p^\dagger c_s\right).
\end{aligned}
\end{equation}
In particular, for $r\ne s$, we move $c_q^\dagger$ to the right in the first term and move $c_q$ to the left in the third term:
\begin{equation}\label{eq:LkCrCs}
\begin{aligned}
    \mathcal{L}_K^\dagger (c_r^\dagger c_s) & =\frac{1}{2}\left(-\sum_{q>s} c_r^\dagger c_s c_q^\dagger c_q-\sum_{p<r} c_p c_p^\dagger  c_r^\dagger c_s -\sum_{q>r} c_q^\dagger c_q c_r^\dagger  c_s -\sum_{p<s}   c_r^\dagger c_s c_p c_p^\dagger\right)\\
    &= -\frac{1}{2} \left(M_r c_r^\dagger c_s + c_r^\dagger c_s M_s\right)
\end{aligned}
\end{equation}
Here, $M_k$ is defined as follows:
\begin{equation}{\label{eq:Mks}}
M_k := \sum_{p<k} c_p c_p^\dagger    +\sum_{q>k} c_q^\dagger c_q = \sum_{p<k} c_p c_p^\dagger +\sum_{q>k} c_q^\dagger c_q = \sum_{p<k} (1-n_p) + \sum_{q>k} n_q.
\end{equation}
Notice that for $r<s$:
\begin{equation}
\begin{aligned}
    \left[c_r^\dagger c_s , M_s\right] &= \sum_{p<s} c_r^\dagger c_s c_p c_p^\dagger  +\sum_{q>s} c_r^\dagger c_s c_q^\dagger c_q -\sum_{p<s} c_p c_p^\dagger c_r^\dagger c_s   -\sum_{q>s}c_q^\dagger c_q  c_r^\dagger c_s  \\
 &= \sum_{p<s} \delta_{pr} c_p^\dagger c_s-\sum_{q>s}\delta_{qr} c_q^\dagger c_s     =     c_r^\dagger c_s . 
    \end{aligned}
\end{equation}
Similarly, for $r>s$:
\begin{equation}\label{eq:r>s}
\begin{aligned}
    [M_r, c_r^\dagger c_s] &=  \sum_{p<r} c_p c_p^\dagger c_r^\dagger c_s + \sum_{q>r} c_q^\dagger c_q c_r^\dagger c_s -\sum_{p<r} c_r^\dagger c_s c_p c_p^\dagger  -\sum_{q>r}  c_r^\dagger c_s c_q^\dagger c_q\\
    &= \sum_{p<r} \delta_{ps} c_r^\dagger c_p  -\sum_{q>r} \delta_{qs}  c_r^\dagger c_q =  c_r^\dagger c_s.
    \end{aligned}
\end{equation}

Using Eq. (\ref{eq:coh}) and Eq. (\ref{eq:LkCrCs}) to (\ref{eq:r>s}), we obtain the equation of motion of the 1-RDM in the molecular orbital basis $\widetilde{P} = \Phi^\dagger P \Phi =\left( \Tr(\rho c_j^\dagger c_i)\right)_{1\le i,j \le 2L}$. For the off-diagonal entries, we have
\begin{equation}\label{eq:offdiag1rdmtype2}
\partial_t \widetilde{P}_{sr} =\begin{cases}
    -\mathrm{i}(\varepsilon_s-\varepsilon_r)\widetilde P_{sr}-\frac{1}{2} \langle( M_r+M_s+1)c_r^\dagger c_s \rangle & r<s,\\
     -\mathrm{i}(\varepsilon_s-\varepsilon_r)\widetilde P_{sr}-\frac{1}{2} \langle c_r^\dagger c_s( M_r+M_s+1) \rangle & r>s.
\end{cases} 
\end{equation}
For the diagonal entries, we have
\begin{equation}\label{eq:diag1rdmtype2}
\begin{aligned}
    \partial_t \widetilde P_{rr} &= \sum_{q>r}\langle c_q^\dagger c_r c_r^\dagger c_q\rangle  - \sum_{p<r} \langle c_r^\dagger c_p c_p^\dagger c_r \rangle = \sum_{q>r } (\widetilde{P}_{qq} - \langle n_r n_q \rangle )+ \sum_{p<r}( \widetilde{P}_{pp} - \langle n_p n_r \rangle)\\
    &= -\sum_{p<r}\ev{(1-n_p)n_r} + \sum_{q>r }\ev{(1-n_r) n_q} = -\langle M_r n_r \rangle  +\sum_{q>r}\langle n_q\rangle.
    \end{aligned}
 \end{equation}
\section{Convergence of 1-RDM in the simplified Hartree-Fock setting}\label{sec:convergence}

For the Type-I setting, we will show the universal exponential convergence that is oblivious to specific chemical details. For the Type-II setting, we provide a qualitative analysis on the convergence rate of the entries of the 1-RDM via mean-field approximation.

As discussed in the main text, the dynamics with Type-I set will converge exponentially to the stationary point $P^\star$ and the mixing time is independent of the system size:

\begin{equation}\label{eq:exps}
  \norm{P(t) -P'(t)}_{\rm F} = e^{-t}\norm{P(0)-P'(0)}_{\rm F}.
\end{equation}

We can easily see Eq. (\ref{eq:exps}) by vectorizing $P-P'$, i.e. let $X:=\mathrm{vec}(P-P')$, then the equation of motion for $X(t)$ takes the following homogeneous form:
\begin{equation}
  \dot{X}(t) = -\mathrm{i}\mathbf{F} X(t) - X(t)\quad \Rightarrow \quad X(t) =e^{-t} e^{-\mathrm{i}\mathbf{F}t } X(0).
\end{equation}
Since $\mathbf{F} = I\otimes F - F^T\otimes I$ is Hermitian, it follows readily that
  \begin{equation}
       \norm{P(t)-P'(t)}_\mathrm{F}  = \norm{X(t)}_2  = e^{-t} \norm{e^{-\mathrm{i}\mathbf{F}t } X(0)}_2  = e^{-t}\norm{X(0)}_2 = e^{-t} \norm{P(0)-P'(0)}_\mathrm{F}.
    \end{equation}
If we define the mixing time of the quasi-free Lindblad dynamics as \cite{DingChenLin2024}
\begin{equation}
 \norm{P(t_{\rm mix}) - P'(t_{\rm mix})}_\mathrm{F} \le \frac{1}{2}\norm{P(0) - P'(0)}_\mathrm{F}
\end{equation}
for any initial covariance matrices $P(0)$ and $P'(0)$, then the dynamics of HF state preparation has the minimum mixing time $\log 2$ by definition and is independent of system size. 

For the Type-II setting, we first estimate the bound of the positive semidefinite matrices $M_k$, defined in Eq. (\ref{eq:Mks}). Let us consider three cases:
\begin{enumerate}
  \item   When $k<N_{\rm e}$, we have $\sum_{q>k }n_q \ge N_{\rm e}-k-1 \ge 1$, therefore $M_k\succeq 1$.
  \item When $k>N_{\rm e}+1$, we have $\sum_{p<k} (1-n_p) \ge k-N_{\rm e} \ge 1$, therefore $M_k\succeq 1$.
  \item For $k\in \{N_{\rm e}, N_{\rm e}+1\}$, we do not have 
 a nontrivial lower bound for $M_k$. Indeed, $M_k$ is positive semidefinite but not positive definite in this case, since we can easily verify that the Hartree-Fock state $\ket{\rm HF}\in \ker M_k$.
\end{enumerate}

For any 2-RDM $\langle a_i^\dagger a_j a_k^\dagger a_\ell\rangle$, we adopt the following mean-field approximation
\begin{equation}
    \langle a_i^\dagger a_j a_k^\dagger a_\ell \rangle \approx \langle a_i^\dagger a_j\rangle \langle a_k^\dagger a_\ell\rangle.
\end{equation}
To qualitatively understand the convergence behavior, we consider the mean-field approximation of Eq. (\ref{eq:offdiag1rdmtype2}) and Eq.(\ref{eq:diag1rdmtype2}). We begin with the off-diagonal entries. For both $r>s$ and $r<s$, we have
\begin{equation}
    \partial_t \widetilde{P}_{sr} \approx \left[  -\frac{1}{2}(1+ \langle M_r+M_s\rangle) + \mathrm{i}(\varepsilon_r -\varepsilon_s)\right] \widetilde{P}_{sr}.
\end{equation}
Noticing from Eq. (\ref{eq:Mks}) that each $M_k$ is a diagonal matrix in the energy basis represented as the sum of two positive semidefinite matrices, we have $M_r + M_s \succeq 0$. It follows readily that under the mean-field approximation, the off-diagonal entry $\widetilde{P}_{sr}$ converges exponentially to zero with an exponent of at least $\frac{1}{2}$.

For the diagonal entries, we take the mean-field approximation on Eq. (\ref{eq:diag1rdmtype2}), yielding
\begin{equation}
    \partial_t \widetilde{P}_{rr} \approx  -\langle M_r\rangle \widetilde{P}_{rr} +\sum_{q>r} \widetilde{P}_{qq}.
\end{equation}
These equations can be solved one-by-one. Specifically, we start from $r=2L$:
\begin{equation}
    \partial_t \widetilde{P}_{2L,2L} \approx -\langle M_{2L}\rangle  \widetilde{P}_{2L,2L}.
\end{equation}
This homogeneous equation gives that $\widetilde{P}_{2L, 2L}$ converges exponentially with an exponent being at least $1$ since $M_{2L}\succeq 1$. Then we move on to $r=2L-1$ and this time $\widetilde{P}_{2L,2L}$ becomes the inhomogeneous term, so the convergence exponent of $\widetilde{P}_{2L-1,2L-1}$ is also greater than $1$ and so forth. This argument holds for all $r>N_{\rm e}+1$. For $r<N_{\rm e}$, we rewrite the mean-field equation of motion as follows:
\begin{equation}
    \partial_t (1-\widetilde{P}_{rr}) \approx \langle M_r\rangle (1- \widetilde{P}_{rr} ) +\sum_{p<r} (1-\widetilde{P}_{pp}).
\end{equation}
We solve these equations from $r=1$ and similarly we conclude that $\widetilde{P}_{rr}$ converges exponentially to $1$ for all $r<N_{\rm e}$.

\section{Proof of Theorem 1 for the spectral gap of the Lindbladian}\label{sec:proofthm1}
\begin{proof}
Since $[\hat{H},\hat{H}_{\rm dp}]=0$, we can choose $\{\ket{\psi_j}\}$ to diagonalize $\hat{H}$ and $\hat{H}_{\rm dp}$ simultaneously, and the eigenvalues for $\hat{H}$ still satisfies $\lambda_0<\lambda_1\le \ldots$. We can also write $\hat{H}_{\rm dp}\ket{\psi_j}=\xi_{j}\ket{\psi_j}$, but the eigenvalues $\xi_{j}$ may not be ordered. 
Then for $\hat{J}=-\mathrm{i}\hat{H}-\hat{H}_{\rm dp}$
we have $\hat J\ket{\psi_j}=(-\mathrm{i}\lambda_j-\xi_{j}) \ket{\psi_j}$. 

We can then use $\{\ketbra{\psi_i}{\psi_j}\}$ as an ordered basis for expanding $\rho$, following a row-major order. Since $\hat{K}_k$ is by construction an upper triangular matrix in the energy basis, 
the operator $\mathcal{K}:\rho \mapsto \sum_k\hat{K}_k \rho \hat{K}_k^{\dag}$ is also upper triangular after vectorization. Meanwhile for $\rho=\ketbra{\psi_i}{\psi_j}$, we have
\begin{equation}
\hat{J}\rho+\rho \hat{J}^{\dag}=(-\mathrm{i}\lambda_i-\xi_{i}+\mathrm{i} \lambda_j-\xi_{j}) \rho.
\end{equation}
So the vectorized Lindbladian, denoted by $\mathbf{L}$, is an upper triangular matrix with eigenvalues $(-\mathrm{i}\lambda_i-\xi_{i}+\mathrm{i} \lambda_j-\xi_{j})$. Since $\hat{H}_{\rm dp}$ has an eigenvalue $0$, this immediately shows that the spectral gap of $\mathbf{L}$ is equal to that of $\hat{H}_{\rm dp}$. 
\end{proof}

We compute the jump operators with Type-I set under the molecular orbital basis. Recall that
\begin{equation}\label{eq:atombasis}
    \hat{K}_{p,+} =  \sum_{r=1}^{2L} a_r^\dagger (\hat{f}(F))_{r,p},\quad \hat K_{q,-} = \sum_{r=1}^{2L} a_r (\hat {f}(-F))_{q,r}.
\end{equation}
Rewriting Eq. (\ref{eq:atombasis}) in the molecular orbital basis yields
\begin{equation}
    \hat{K}_{p,+}= \sum_{r=1}^{2L} c_r^\dagger \hat{f}(\varepsilon_r) \Phi_{pr}^\ast,\quad \hat{K}_{q,-} = \sum_{r=1}^{2L} c_r \hat{f}(-\varepsilon_r) \Phi_{qr}.
\end{equation}
Then
\begin{equation}
    \hat H_{\rm dp} = \frac{1}{2}\sum_{p} \hat K_{p,+}^\dagger \hat K_{p,+} + \frac{1}{2}\sum_{q} \hat K_{q,-}^\dagger \hat K_{q,-} = \frac{1}{2}\sum_{r=1}^{2L} \hat{f}^2(\varepsilon_r) c_rc_r^\dagger + \frac{1}{2}\sum_{r=1}^{2L} \hat{f}^2(-\varepsilon_r) c_r^\dagger c_r.
\end{equation}
In other words,
\begin{equation}
    \hat{H}_{\rm dp} = \frac{1}{2}\sum_{p\le N_{\rm e}}(1-n_p) + \frac{1}{2}\sum_{q>N_{\rm e}} n_q.
\end{equation}

\section{Supplementary notes on numerical simulation results for \emph{ab initio} calculations}\label{sec:notesOnNumerics}

To construct the converged HF Hamiltonian in second quantized representation for the numerical validations of the convergence rate arguments of the simplified HF settings (with Type-I or Type-II set), we first apply the L\"owdin's orthogonalization to obtain a set of orthonormal spin orbitals \cite{Loewdin1950, SzaboOstlund2012}. Concretely, given that the overlap matrix $S$ is Hermitian positive definite, we can diagonalize it as $S= UsU^\dagger$ where $U$ is a unitary matrix and $s$ is a diagonal matrix of positive eigenvalues. We define the transformation matrix $X = Us^{-1/2}$ and transform the Fock operator using $F\mapsto X^\dagger F X$. This transformation is equivalent to using the orthogonalized spin orbitals as the new basis set. The one-particle and two-particle integrals, as well as the overlap integrals within a given basis set are imported from the \texttt{PySCF} package \cite{SunBerkelbachBluntEtAl2018}. 

We can then consider the electronic structure Hamiltonian in the FCI basis
\begin{equation}
  \hat H = \sum_{p,q=1}^{2L} T_{pq}c_p^\dagger c_q +\frac{1}{2}\sum_{p,q,r,s=1}^{2L} S_{pqrs} c_p^\dagger c_q^\dagger c_r c_s,
\end{equation}
where the $T_{pq}$ and $S_{pqrs}$ are the one-particle and two-particle integrals of molecular orbitals generated from the HF calculation.

For all of the simulations, we choose the parameters of the filter as follows: $a= 2.5\lVert \hat H\rVert$, $\delta_a = a/5$, $b=\delta_b =\Delta$. For the integral truncation, we let $S_s= \frac{10}{\Delta}$ and for the quadrature nodes we set $M_s = \lceil S_s/(\pi/2a)\rceil $ and $\Delta s = S_s/M_s$ \cite{DingChenLin2024}. All units are in the atomic unit (a.u.).

We numerically verify that this method can indeed approximate Lindblad dynamics. We use the vacuum state as initial state and the reduced Type-I set $\mathcal{S}_{\rm I}$ as the jump operators to prepare the ground state of $\rm H_2$ system. We apply the quantum jump unraveling method for these simulations with time step $\Delta t = 0.1$ and the stopping times are set to $T=30$. In each simulation, we repeat the pure-state evolution on $N_{\rm traj} = 10, 25, 50, 100$ and $500$ trajectories and we approximate the many-body density operator at the time $t_n$ as:
\begin{equation}
  \rho (t_n) =\mathbb{E}\dyad{\psi(t_n)} \approx\frac{1}{N_{\rm traj}} \sum_{k=1}^{N_{\rm traj}} \dyad{\psi_k(t_n)},
\end{equation}
where $\ket{\psi_k(t_n)}$ denotes the state vector on the $k$-th trajectory at the time $t_n$. It can observed in Supplementary Figure \ref{fig:trajs} that as the number of trajectories increases, the results of the quantum-jump unraveling exhibit a trend of convergence to the exact Lindblad dynamics, which is obtained by propagating the density operator using the DOPRI5 solver.

\begin{figure}[htbp]
    \centering
\includegraphics[width=0.95\linewidth]{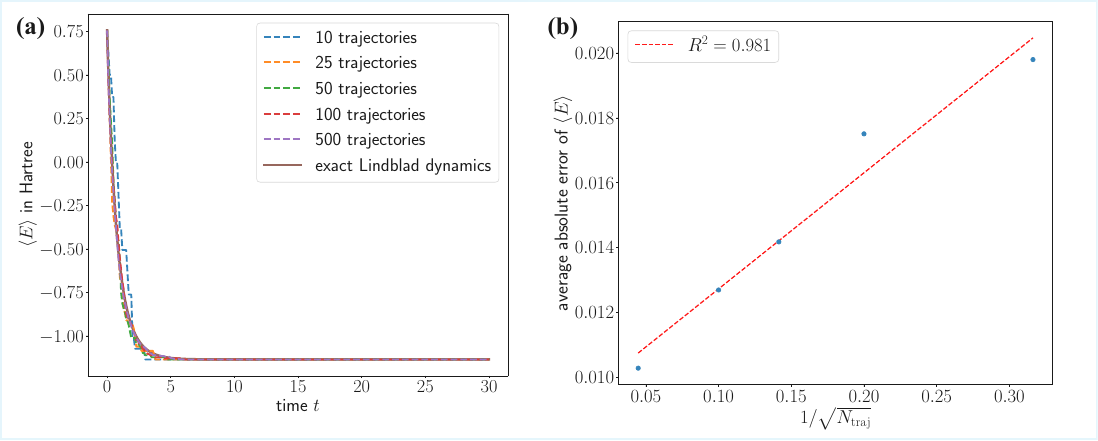}
    \caption{\textbf{The Lindblad dynamics of the $\rm H_2$ system in the STO-3G    basis set, obtained by exactly propagating the many-body density operator and using the quantum-jump unraveling method with 10, 25, 50, 100, and 500 trajectories, respectively.} The \textbf{(a)} panel illustrates the energy convergence curves of each simulation. The \textbf{(b)} panel shows the average absolute error of the overlap $\langle \rho_{g}\rangle$ with the exact ground state relative to the exact Lindblad dynamics at each time point. The average error decreases approximately as $1/\sqrt{N_{\mathrm{traj}}}$. }
    \label{fig:trajs}
\end{figure}

Due to the quadrature error in the jump operators, the Monte Carlo trajectory evolution methods may exhibit numerical instability. In short, this instability arises because the quadrature error can occasionally lead to an overestimation of the quantum jump probability, which can cause an increase in energy, particularly as the dynamics approach the ground state.
\section{Molecular geometries used for the numerical simulations}

For the numerical simulations of the ground state preparation at the Hartree-Fock (HF) level using the quasi-free approach, we conduct the simulation for these molecular systems: $\rm H_4,$ $\rm LiH$, $\rm H_2O$, $\rm CH_4$, $\rm HCN$, $\rm C_2H_4$, $\rm N_2$, $\rm H_{10}$ and $\rm SO_3$. The molecular geometries are listed below:

\noindent$\mathbf{H_4}$ in STO-3G:\\
\noindent\texttt{
H 0.0 0.0 0.0\\
H 0.0 0.0 2.0\\
H 0.0 2.0 0.0\\
H 0.0 2.0 2.0\\
}

\noindent$\mathbf{LiH}$ in STO-3G:\\
\noindent\texttt{
H 0.000 0.000 0.000\\
Li 0.000 0.000 1.546\\
}

\noindent$\mathbf{H_2O}$ in STO-3G:\\
\noindent\texttt{
O       0.0000000       0.0000000       0.1271610\\
H       0.0000000       0.7580820       -0.5086420\\
H       0.0000000       -0.7580820      -0.5086420\\
}

\noindent$\mathbf{HCN}$ in STO-3G:\\
\noindent\texttt{
C       0.0000000       0.0000000       -0.5000780\\
H       0.0000000       0.0000000       -1.5699900\\
N       0.0000000       0.0000000       0.6529230\\
}

\noindent$\mathbf{C_2H_4}$ in STO-3G:\\
\noindent\texttt{
C       0.0000000       0.0000000       0.6530360\\
C       0.0000000       0.0000000       -0.6530360\\
H       0.0000000       0.9157470       1.2292130\\
H       0.0000000       -0.9157470      1.2292130\\
H       0.0000000       -0.9157470      -1.2292130\\
H       0.0000000       0.9157470       -1.2292130\\
}

\noindent$\mathbf{N_2}$ in cc-pVDZ:\\
\noindent\texttt{
N       0.0000000       0.0000000       0.5386530\\
N       0.0000000       0.0000000       -0.5386530\\
}

\noindent$\mathbf{H_{10}}$ in cc-pVDZ:\\
\noindent\texttt{
H       0.0     0.0     0.0\\
H       0.0     0.0     0.7\\
H       0.0     0.0     1.4\\
H       0.0     0.0     2.1\\
H       0.0     0.0     2.8\\
H       0.0     0.0     3.5\\
H       0.0     0.0     4.2\\
H       0.0     0.0     4.9\\
H       0.0     0.0     5.6\\
H       0.0     0.0     6.3\\
}

\noindent$\mathbf{SO_3}$ in cc-pVDZ:\\
\noindent\texttt{
S       0.0000000       0.0000000       0.0000000\\
O       0.0000000       1.4167620       0.0000000\\
O       1.2269520       -0.7083810      0.0000000\\
O       -1.2269520      -0.7083810      0.0000000\\
}

\begin{table}[htbp]
    \centering
 \caption{\label{tab:geo1}The molecular geometries used in the simulations of interacting systems with Type-I}

    \begin{tabular}{ccccc}
    \toprule
     molecule & basis set   &  bond length (\AA)&  number of qubits & dimension of $\mathcal{F}$ \\
     \midrule
        $\rm H_2$ & STO-3G& 0.7 &  4 & 16\\
        $\rm H_2$ & 6-31G&  0.7 &  8 &  256 \\
        $\rm H_4$ (chain)& STO-3G & 0.7 & 8& 256\\
        $\rm H_6$ (chain) & STO-3G & 0.7 &12& 4096\\
             \bottomrule
    \end{tabular}

\end{table}

\begin{table}[htbp]
   \centering
    \caption{\label{tab:geo2}The molecular geometries used in the simulations of interacting systems with Type-II}
 
    \begin{tabular}{ccccc}
    \toprule
     molecule & basis set   &  bond length (\AA)&  number of qubits & dimension of $\mathcal{F}_{N_{\rm e}}$ \\
     \midrule
        $\rm H_4$ (square) & STO-3G & - & 8 & 70\\
        $\rm H_2$ & cc-pVDZ & 0.7 & 20 & 190\\
        $\rm F_2$ & STO-3G & 1.4 & 20 & 190\\
        $\rm LiH$ & STO-3G & 1.546 & 12 & 495\\
        $\rm Cl_2$ & STO-3G & 2.130 & 36 & 630\\
        $\rm H_2O$ & STO-3G & - & 14 & 1001\\
        $\rm BeH_2$ & STO-3G & 1.304 & 14 & 3003\\
            \bottomrule
    \end{tabular}

\end{table}

For the simulations of interacting systems using the Monte Carlo trajectory-based method, we employed the same geometry as in the quasi-free calculation for the $\rm H_2O$ molecule within the STO-3G basis set. The geometries for other molecular systems are listed in Supplementary Table \ref{tab:geo1} and Supplementary Table \ref{tab:geo2}. Additionally, we provide the dimension of the total Fock space $\mathcal{F}$ and the FCI sector $\mathcal{F}_{N_{\rm e}}$ with $N_{\rm e}$ electrons, which corresponds to the size of the Hamiltonian matrix $\hat{H}$. 
\vspace{1cm}

\bibliographystyle{naturemag} 

\end{document}